\documentclass[fleqn,usenatbib]{mnras}

\usepackage[T1]{fontenc}
\usepackage{ae,aecompl}

\def\surfs{{\sc surfs}}
\def\shark{{\sc Shark}}
\def\viperfish{{\sc Viperfish}}
\def\prospect{{\sc ProSpect}}

\usepackage{graphicx}	
\usepackage{amsmath}	
\usepackage{amssymb}	

\title[Physical properties of SMGs in \shark]{Physical properties and evolution of (Sub-)millimeter selected galaxies in the galaxy formation simulation \shark}

\author[Claudia del P. Lagos et al.]{Claudia del P. Lagos$^{1,2,3}$\thanks{E-mail: claudia.lagos@icrar.org}, 
Elisabete da Cunha$^{1,2}$, Aaron S.~G. Robotham$^{1,2}$
\newauthor{Danail Obreschkow$^{1,2}$, Francesco Valentino$^{3,4}$, Seiji Fujimoto$^{3,4}$,}
\newauthor{Georgios E. Magdis$^{3,4,5}$, Rodrigo Tobar$^{1}$}
\\
$^{1}$International Centre for Radio Astronomy Research (ICRAR), M468, University of Western Australia, 35 Stirling Hwy, Crawley, \\WA 6009, Australia.\\
$^{2}$ARC Centre of Excellence for All Sky Astrophysics in 3 Dimensions (ASTRO 3D).\\
$^{3}$Cosmic Dawn Center (DAWN).\\
$^{4}$Niels Bohr Institute, University of Copenhagen, Lyngbyvej 2, DK-2100 Copenhagen, Denmark.\\
$^{5}$DTU-Space, Technical University of Denmark, Elektrovej 327, DK-2800 Kgs. Lyngby, Denmark. 
}

\date{Accepted XXX. Received YYY; in original form ZZZ}

\pubyear{2020}

\begin{document}
\label{firstpage}
\pagerange{\pageref{firstpage}--\pageref{lastpage}}
\maketitle

\begin{abstract}
We thoroughly explore the properties of (sub)-millimeter (mm) selected galaxies (SMGs) in the \shark\ semi-analytic model of galaxy formation.
Compared to observations, the predicted number counts at wavelengths ($\lambda$) $0.6-2$mm and redshift distributions at $0.1-2$mm, agree well. At the bright end ($\gtrsim 1$~mJy), \shark\ galaxies are a mix of mergers and disk instabilities. These galaxies display a stacked FUV-to-FIR spectrum that agrees well with observations. We predict that current optical/NIR surveys are deep enough to detect bright ($>1$~mJy) $\lambda=0.85-2$mm-selected galaxies at $z\lesssim 5$, but too shallow to detect counterparts at higher redshift. A James Webb Space Telescope 10,000s survey should detect {\it all} counterparts for galaxies with $S_{\rm 0.85mm}\gtrsim 0.01$~mJy. We predict SMG's disks contribute significantly (negligibly) to the rest-frame UV (IR). We investigate the $0\le z\le 6$ evolution of the intrinsic properties of $>1$~mJy $\lambda=0.85-2$mm-selected galaxies finding their: (i) stellar masses are $>10^{10.2}\rm\, M_{\odot}$, with the $2$mm ones tracing the most massive galaxies ($>10^{11}\rm\, M_{\odot}$); (ii) specific star formation rates (SFR) are mildly ($\approx 3-10\times$) above the main sequence (MS); (iii) host halo masses are $\gtrsim 10^{12.3}\,\rm M_{\odot}$, with $2$mm galaxies tracing the most massive halos (proto-clusters); (iv) SMGs have lower dust masses ($\approx 10^{8}\,\rm M_{\odot}$), higher dust temperatures ($\approx 40-45$~K) and higher rest-frame V-band attenuation ($>1.5$) than MS galaxies; (v) sizes decrease with redshift, from $4$~kpc at $z=1$ to $\lesssim 1$~kpc at $z=4$; (vi) the Carbon Monoxide line spectra of $S_{\rm 0.85mm}\gtrsim 1$~mJy sources peak at $4\rightarrow 3$. Finally, we study the contribution of SMGs to the molecular gas and cosmic SFR density at $0\le z\le 10$, finding that $>1$~mJy sources make a negligible contribution at $z\gtrsim 3$ and $z\gtrsim 5$, respectively, suggesting current observations have unveiled the majority of the star formation at $0\le z\le 10$.
\end{abstract}




\section{Introduction}

One of the most intriguing galaxy populations observed to date are the so-called submillimeter galaxies (SMGs) - those that appear bright in the submillimeter (submm) or millimeter (mm) bands (typically with fluxes $\gtrsim 1$~mJy). They are rare in the local Universe, but are increasingly common with lookback time, with their number density peaking at around $z\approx 2$ \citep{Chapman05,Casey14}. \citet{Casey12} showed that these galaxies make a negligible contribution to the total cosmic star formation rate density (CSFRD) locally, but make up $\gtrsim 50$\% of the CSFRD at $z\gtrsim 1$. Given their large contribution to the CSFRD in the early Universe and their number density, they are thought to be the progenitors of local massive elliptical galaxies (e.g. \citealt{Cai13,Toft14,Valentino20}). 
Hence, understanding the nature of SMGs and more generally, submm and mm-selected galaxies, is fundamental to unveiling the formation of galaxies at cosmic noon. Throughout this paper we will loosely refer to SMGs as galaxies that were selected from their submm or mm emission.

The advent of the Atacama Large Millimeter Array (ALMA) has opened an unprecedented window to study SMGs at high angular resolution and sensitivity across cosmic time. \citet{Karim13}, using ALMA, showed that the number counts of bright SMGs (those with fluxes at the observer-frame $870\mu$m $\gtrsim 1$~mJy) obtained with previous single dish telescopes were highly affected by confusion, offering the first accurate measurements of the abundance of bright galaxies at $870\mu$m. ALMA has also allowed measurements of the number counts to go much deeper than ever before. \citet{Fujimoto16,GonzalezLopez20} presented $1$~mm number counts going as deep as $0.01$~mJy and $0.03$~mJy, respectively, {using ALMA} with the former finding $\approx 2.5$ times more faint sources than the latter {(see also \citealt{Munoz-Arancibia18})}. This difference is significant and implies that {$S>0.01$~mJy sources recover the full far-infrared (FIR) background light or there is significant emission from fainter sources.}
It is however unclear whether these differences are solely due to cosmic variance or there are additional effects in play. {The former is particularly significant given the pencil-beam surveys carried out so far with ALMA.}
ALMA has also allowed to remeasured the redshift distribution for these bright SMGs using both photometric and spectroscopic redshift determinations, showing that their abundance is highest at $z\approx 2-2.7$ \citep{daCunha15,Dudzeviciute20}. Selecting on longer wavelengths generally leads to a higher redshift peak \citep{Reuter20}. Despite this progress, most SMG redshifts come from their multi-wavelength photometry, which can suffer from significant systematic effects (\citealt{Battisti19}; e.g. $44$ out of 
$707$ galaxies in \citealt{Dudzeviciute20} have confirmed redshifts). Spectroscopic redshift measurements require spectral scan techniques that can be time consuming but are nonetheless an important quest.

ALMA has also offered the opportunity to study the morphology of SMGs, overall finding that about half of the bright $\gtrsim 1$~mJy $870\mu$m sources have disturbed morphologies (such as tidal tails or multiple nuclei), indicative of being galaxy merger-driven, while the other half are consistent with being disks (e.g. \citealt{Cowie18,Hodge19,Gullberg19}). This is dramatically different to the local Universe, in which all infrared (IR)-bright galaxies have morphologies consistent with being ongoing galaxy mergers \citep{Veilleux02,daCunha10}. This shows that these high-z SMGs are an inhomogeneous galaxy population. Besides their morphology, intrinsic properties, such as stellar masses and star formation rates (SFRs) have been derived, showing that bright SMGs are massive galaxies, $M_{\star}\approx 10^{10}-10^{11}\,\rm M_{\odot}$, with SFRs $\approx 200-300\,\rm M_{\odot}$, on average \citep{daCunha15,Dudzeviciute20}. This indicates that these SMGs are only modestly above the star-forming main sequence (MS) in the SFR-stellar mass plane, by factors of $\approx 3-5$, on average. This again shows the different nature of these high-z SMGs compared to local ones, which are associated with extreme starbursts that are $\gtrsim 10$ times above the local Universe MS. This is not necessarily surprising, as the MS is strongly evolving with redshift, and hence high SFRs are more common in the early Universe \citep{Elbaz10,Whitaker12}.
Interestingly, measurements of the sizes of these bright SMGs show that they are not so different from MS galaxies of the same stellar mass \citep{Ikarashi15,Barro16,Simpson15,Fujimoto17,Puglisi19}.

Despite this tremendous progress there are still areas that require significant exploration. \citet{Casey18a} highlighted the fact that the very limited survey areas that have been accessed with ALMA imply that we still do not have firm constraints on the contribution of SMGs to the CSFRD of the Universe at $z\gtrsim 4$. \citet{Casey18a} showed that in some extreme models of dusty galaxies this contribution may be several times higher than that measured from rest-frame ultraviolet (UV)-selected galaxies \citep{Bouwens12}. When studying the multi-wavelength properties of SMGs, it is also clear that a significant fraction do not have optical-to-near IR (NIR) counterparts. \citet{Dudzeviciute20} found that $10$\% of their $707$ SMGs do not have detections in the IR Array Camera (IRAC) on the Spitzer Space Telescope, which either imply much higher redshifts or extreme obscuration. The upcoming James Webb Space Telescope (JWST) promises to change the landscape again, providing detections for most if not all observed SMGs to date {as we show in this paper}.

In galaxy formation simulations, the abundance and properties of SMGs have been notoriously difficult to reproduce \citep{Casey14}. {This has been investigated using both semi-analytic models and hydrodynamical simulations of galaxy formation. The former have the advantage of accessing much larger cosmological volumes allowing for robust statistical studies of the abundance of SMGs at different redshifts, at the expense of the detailed internal structure predictions of galaxies. On the other hand, the advantage of hydrodynamical simulations resides in predicting the internal structure of gas in galaxies, from which one can link to the morphologies being studied with ALMA, at the expense of sample size. These two techniques are therefore highly complementary.
Regarding SMGs,}
\citet{Baugh05}, using the GALFORM semi-analytic model of galaxy formation, showed that assuming a universal initial stellar mass function (IMF) of stars made the model incapable of reproducing their number counts and redshift distribution, suggesting that a plausible solution was to allow for a top-heavy IMF during starbursts. This was latter confirmed by \citet{Lacey15} with updated versions of GALFORM. \citet{Hayward11,Hayward13} combined $34$ idealised isolated galaxies and galaxy mergers hydrodynamical simulations with various empirical cosmological relations to get a statistical estimate of number counts and redshift distributions. They argued that variations to the IMF were not necessary to reproduce these observations; however, \citet{Cowley19} showed that the assumptions for how to populate large galaxy samples with $850\mu$m proposed by \citet{Hayward11} were inconsistent with the self-consistent predictions of GALFORM obtained using the radiative transfer (RT) code {\sc GRASIL} \citep{Granato00}. The latter predicted $\lesssim 10$ times less bright SMGs than the ones one would derived from applying the \citet{Hayward11} prescriptions, showing that self-consistent predictions cannot be replaced by semi-empirical relations.

Significant progress has been recently reported in the literature in this area. \citet{Lagos19},  using the \shark\ semi-analytic model of galaxy formation \citep{Lagos18b} in combination with the spectral energy distribution (SED) code \prospect\ \citep{Robotham20} and the RT analysis of the {\sc EAGLE} hydrodynamical simulations of \citet{Trayford19} showed for the first time that a cosmological galaxy formation simulation was capable of reproducing the far UV-to far IR emission of galaxies in a wide redshift range, including the SMG number counts and redshift distributions, in a self-consistent way using a universal IMF. The key to the success of \shark\ was attributed to the adoption of attenuation curves that scale with the dust surface density of galaxies as reported by \citet{Trayford19}, which take into account the 3D distribution of star formation and dust in galaxies. \citet{Lovell20} using the {\sc Simba} hydrodynamical simulations \citep{Dave19} coupled with the RT code {\sc Powerday} \citep{Narayanan20} also found that their model was able to reproduce the SMG number counts and redshift distributions using a universal IMF. The fact that two independent galaxy formation simulations reached the same conclusions, robustly shows that a variable IMF is not required to reproduce these key observations of SMGs.

In the area of SMG intrinsic properties, \citet{McAlpine19} presented a thorough study of SMGs at $1< z< 3$ with $850\mu$m fluxes $>1$~mJy in the {\sc EAGLE} simulations \citep{Schaye14,Crain15}. They found these high-z SMGs are massive, $M_{\star}\approx 10^{11}\,\rm M_{\odot}$, and have SFR~$\approx 100\,\rm M_{\odot}\,yr^{-1}$, in broad agreement with observations. They also find that the incidence of galaxy mergers in bright SMGs is very similar to the underlying galaxy population, suggesting that mergers are not the sole driver of SMGs in {\sc EAGLE}, again in broad agreement with the observational inferences from the morphologies of SMGs. An important drawback of this investigation is that {\sc EAGLE} underpredicts the abundance of bright SMGs \citep{Cowley19} and predicts a significant number of those to be at $z<1$\footnote{{The redshift distribution for {\sc EAGLE} was obtained by selecting galaxies from their public database \citep{McAlpine15,Camps18} with $S_{\rm 850\mu m}>1$~mJy (not shown here), which shows a peak of sources towards $z=0$.}}, in tension with the observations. Ideally, one would like to explore the intrinsic properties of SMGs in a galaxy formation simulation that broadly reproduces the basic observables of number counts and redshift distributions. This is the focus of this work.

We use the \shark\ semi-analytic model and the SED generation pipeline presented in \citet{Lagos19} to study the FUV-to-MIR and carbon monoxide (CO) emission and intrinsic properties of submm and mm-selected galaxies across cosmic time, $0\le z\le 10$. The starting point is the fact that \shark\ reproduces the $850\mu$m number counts and redshift distributions, but here we extend the testing of \shark\ to the full FIR spectrum. This paper is organised as follows: $\S$~\ref{sec:model} presents a description of \shark, the adopted $N$-body simulations and a summary of how several baryonic physical processes are modelled. $\S$~\ref{sec:model} also presents a summary of how SEDs and CO spectral line energy distributions (SLEDs) are generated. $\S$~\ref{overallstats} compares our predictions with observations of number counts at wavelengths from $650\mu$m to $2$mm, and redshift distributions of galaxies selected at wavelengths from $100\mu$m to $2$mm. $\S$~\ref{SEDPropsSMGs} and $\S$~\ref{intrinsicprops} present the predicted UV-to-MIR emission and intrinsic properties of SMGs, respectively, in \shark\ and compares to observations where possible. $\S$~\ref{sec:CSFRD} presents an analysis of the contribution of submm and mm-selected galaxies to the CSFRD and the cosmic molecular gas density at $0\le z\le 10$, tackling the question of whether we are missing a significant SFR and cold gas source at high redshifts. Finally, $\S$~\ref{conclusions} presents our conclusions.

\section{The \shark\ semi-analytic model}\label{sec:model}

\shark\ is an open source, flexible and highly modular SAM\footnote{\href{https://github.com/ICRAR/shark}{\url{https://github.com/ICRAR/shark}}}, introduced in \citet{Lagos18b}. The model runs over merger trees and halo/subhalo populations computed from the \surfs{} N-body simulations. Below we provide a summary of the \surfs{} suite in $\S$~\ref{SURFS}, then describe the physics included in \shark\ in $\S$~\ref{modeldescription}, and finally the way we compute SEDs and CO SLEDs for our simulated galaxies in $\S$~\ref{SEDs} and $\S$~\ref{COmodel}, respectively.

\subsection{The \surfs{} $N$-body suite}\label{SURFS}

\shark\ uses the \surfs\ suite of N-body, dark-matter (DM) only simulations \citep{Elahi18}, most of which have cubic volumes of $210\,\rm cMpc/h$ on a side, and span a range in particle number,
currently up to $8.5$ billion particles using a $\Lambda$CDM \citet{Planck15} cosmology. These correspond to a total matter, baryon and $\Lambda$ densities of $\Omega_{\rm m}=0.3121$, $\Omega_{\rm b}=0.0491$ and $\Omega_{\rm L}=0.6879$, respectively, with a Hubble parameter of $H_{\rm 0}=h\, 100\,\rm Mpc^{-1}\, km\,s^{-1}$ with $h=0.6751$, scalar spectral index of $n_{\rm s}=0.9653$ and a power spectrum normalization of $\sigma_{\rm 8}=0.8150$. 
All simulations were run with a memory lean version of the {\sc gadget2} code on the Magnus supercomputer at the Pawsey Supercomputing Centre. In this paper, we use the L210N1536 simulation, which has a cosmological volume of $(210\,\rm cMpc/h)^3$, $1536^3$ DM particles with a mass of $2.21\times10^8\,\rm h^{-1}\,M_{\odot}$ and a softening length of 
$4.5\,\rm h^{-1}\,ckpc$. Here, cMpc and ckpc denote comoving Mpc and kpc, respectively. 
\surfs\ produces $200$ snapshots for each simulation {at $z=0-24$}, typically having a time span between snapshots in the range of $\approx 6-80$~Myr. So far \surfs\ consists of $7$ simulations, but this is an ever growing suite.

Merger trees and halo catalogs, which are the basis for \shark\ (and generally any SAM), were constructed using the phase-space finder {\sc VELOCIraptor}\footnote{{\url{https://github.com/icrar/VELOCIraptor-STF/}}} \citep{Elahi19a,Canas18} and the halo merger tree code
{\sc TreeFrog}\footnote{\href{https://github.com/pelahi/TreeFrog}{\url{https://github.com/pelahi/TreeFrog}}}, developed to work on {\sc VELOCIraptor} \citep{Elahi19b}. \citet{Poulton18,Poulton19} show that {\sc TreeFrog}+{\sc VELOCIraptor} lead to very well behaved merger trees, with orbits that
are well reconstructed. \citet{Elahi18b} also show that these orbits reproduce the velocity dispersion vs. halo mass inferred in observations. 
\citet{Canas18} show that the same code can be applied to hydrodynamical simulations to identify galaxies and that the performance of {\sc VELOCIraptor} is superior to space-finders, even in complex merger cases. We refer to \citet{Lagos18b} for more details on how the merger trees and halo catalogs are constructed for \shark, and to \citet{Elahi19a,Elahi19b,Canas18,Poulton18} for more details on the {\sc VELOCIraptor} and {\sc TreeFrog} software.

\subsection{Baryon physics in \shark}\label{modeldescription}

\shark\ includes a large range of  physical processes that are key in shaping galaxy formation and evolution. These are (i) the collapse and merging of DM halos; (ii) the accretion of gas onto halos, which is modulated by the DM accretion rate; (iii) the shock heating and radiative cooling of gas inside DM halos, leading to the
formation of galactic disks via conservation of specific angular momentum of the cooling gas; (iv) star formation (SF) in galaxy disks; (v) stellar feedback from the evolving stellar populations; (vi) chemical enrichment of stars and gas; (vii) the growth via gas accretion and merging of supermassive black holes; (viii) heating by AGN; (ix) photoionization of the intergalactic medium; (x) galaxy mergers
driven by dynamical friction within common DM halos which can
trigger starbursts and the formation and/or growth of spheroids; (xi) collapse of globally unstable disks that also lead to starbursts and the formation and/or growth of bulges (we refer to this mechanism as ``disk instabilities''). In \shark, galaxy mergers are considered to be of {\it any} mass ratio, as long as two galaxies are involved in the collision. Disk instabilities on the other hand, are a secular process, which happens in galaxies whose disks are globally unstable. In this case, the disk is {instantaneously} collapsed into a central overdensity (i.e. forming a bulge or contributing to it), and the inflowing gas triggers a starburst.
\shark\ adopts a universal \citet{Chabrier03} initial mass function (IMF). \citet{Lagos18b} include several different models for gas cooling, AGN,
stellar and photo-ionisation feedback, star formation and dynamical friction timescales. Here, we adopt the default \shark\ model (see models and parameters adopted in \citealt{Lagos18b}; their Table~$2$).

In \shark, SF is computed from the surface density of molecular gas. SF in galaxy disks follow $\Sigma_{\rm SFR}=\Sigma_{\rm mol}/\tau_{\rm mol}$, where $\Sigma_{\rm SFR}$ and $\Sigma_{\rm mol}$ are the SFR and molecular gas surface densities, respectively, while $\tau_{\rm mol}=1\,\rm Gyr$ is the molecular gas depletion timescale. The adopted value comes from observational constraints \citep{Leroy08,Bigiel10} {and is assumed to be constant across cosmic time. The latter is motivated by the Carbon Monoxide (CO) observations of galaxies of \citet{Genzel15} and \citet{Schinnerer16} at $0\lesssim z\lesssim 4$, which show little evolution of $\tau_{\rm mol}$}. In the case of starbursts triggered by galaxy mergers and disk instabilites, the SF model used resembles that for SF in disks but with a shorted depletion timescale,  
  $\Sigma_{\rm SFR}=\Sigma_{\rm mol}/(\tau_{\rm mol}/f_{\rm boost})$, with $f_{\rm boost}=10$. The latter is motivated by observational constraints \citep{Sargent14}. We will refer to these two modes of SF as ``normal'' (SF in disks) and ``burst'' (SF triggered by galaxy mergers and global disk instabilities) modes. The latter is the one that builds bulges in \shark. {Although the transfer of gas during galaxy mergers and disk instabilities to the bulge is instantaneous, the SF episode is slow given the SF law we adopt, which means that during starbursts the molecular gas is consumed typically in a few $100$~Myr.}
  
We numerically solve the differential equations (DEs) of mass, metals and angular momentum exchange between the different baryon reservoirs (see Eqs. $49$-$64$ in \citealt{Lagos18b}), only setting an accuracy to which these equations are solved. The baryon reservoirs in the model are: gas outside halos, hot and cold gas inside halos but outside galaxies, ionised/atomic/molecular gas and stars in disks and bulges in galaxies, and super-massive black holes. Together with intrinsic properties of galaxies at the output times, we also store the star formation history (SFH) and metallicity history (ZFH) of the stars that form at each timestep prior to the output time. These are later used to compute each galaxy's SED. The solving of the set of DEs as well as the storing of SFH and ZFH is done for all galaxy stellar components separately: disks and bulges. In addition, we track the two growth mechanisms of the latter, disk instabilities and galaxy mergers. 
Note that at $z\gtrsim 1$, these bulges are very active and harbour high surface densities of SFR (see \citealt{Lagos19} for examples of the SEDs of passive and active bulges in \shark). This tracking of stellar component allows us to build independent SEDs for galaxy disks, bulge stars that formed via galaxy mergers, and bulge stars that formed via disk instabilities. This is important to understand which galaxy components and types of bulges dominate over different luminosities and cosmic times.

The model parameters of the default \shark\ model were tuned to the $z=0,\,1,\,2$ stellar mass functions (SMFs; \citealt{Wright18}), the $z=0$ black hole-bulge mass relation (\citealt{McConnell13}) and mass-size relations (\citealt{Lange16}). The model also reproduces very well observational results that are independent from those used for the tuning, such as the total neutral, atomic and molecular hydrogen-stellar mass scaling relations at z=0, the CSFRD density evolution at $z\approx 0-4$, the cosmic density evolution of the atomic and molecular hydrogen at $z\lesssim 2$ or higher in the case of the latter, the mass-metallicity relations for the gas and stars, the contribution to the stellar mass by bulges and the SFR-stellar mass relation in the local Universe (see \citealt{Lagos18b} for more details). 
\citet{Davies19} also show that \shark\ reproduces the scatter around the MS of star formation in the SFR-stellar mass plane; \citet{Chauhan19,Chauhan20} show that \shark\ reproduces very well the HI mass and velocity width of galaxies and the HI-halo mass relation observed in the ALFALFA survey; \citet{Amarantidis19} show that the AGN LFs agree well with observations in the X-rays and radio wavelengths; and \citet{Bravo20} show that optical colour distributions and passive fractions of \shark\ galaxies agree reasonably well with GAMA observations. These can be seen as true successes of the model as none of these observations were used in the processes of tuning the free parameters.

\subsection{Spectral Energy Distributions of galaxies in \shark}\label{SEDs}
\begin{figure*}
\begin{center}
\includegraphics[trim=35mm 0mm 32mm 10mm, clip,width=0.99\textwidth]{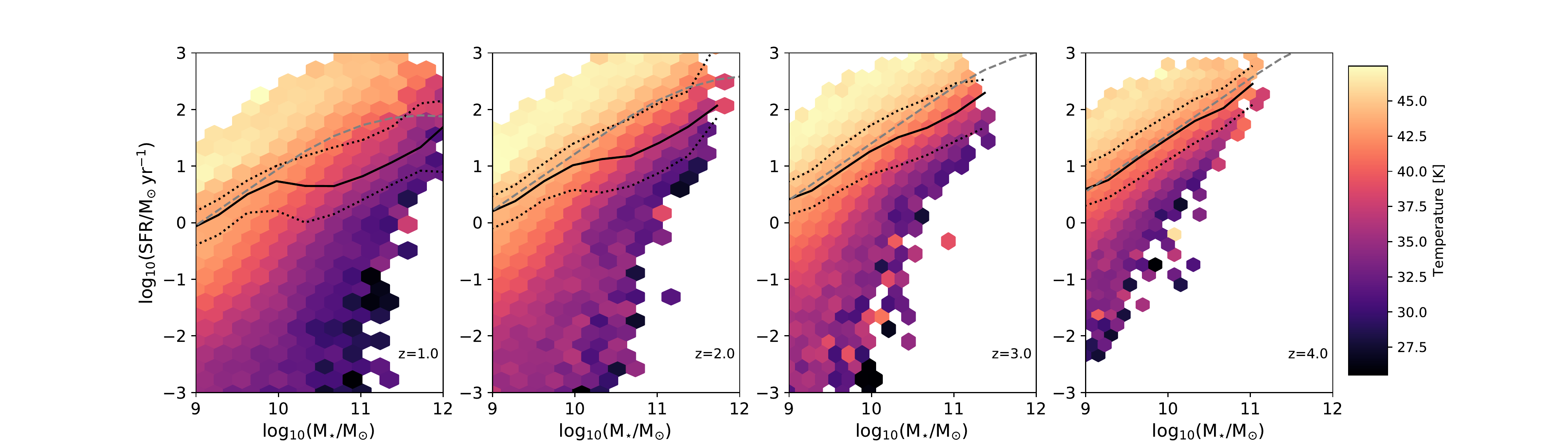}
\includegraphics[trim=35mm 0mm 32mm 10mm, clip,width=0.99\textwidth]{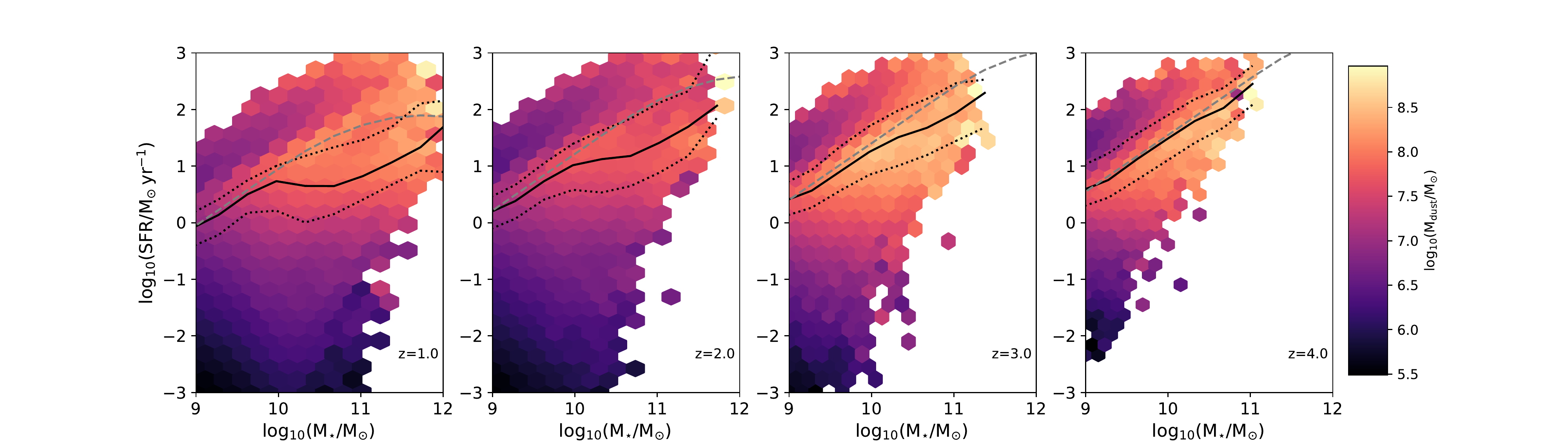}
\caption{{\it Top panels:} SFR vs. stellar masses plane at four different redshifts from $z=1$ to $z=4$, as labelled, using a $\Delta z=0.15$. Pixels with $\ge 10$ galaxies are coloured by their median dust temperature. Solid and dotted lines show the median SFR and $16^{\rm th}-84^{\rm th}$ percentile ranges in bins of stellar mass, respectively, of all galaxies with $\rm SFR>0$. This can be considered as the MS in \shark, except at the high-mass end ($\gtrsim 10^{10.2}\,\rm M_{\odot}$), where galaxies are affected by AGN feedback at $z\lesssim 3$. For reference, we also show the MS inferred observationally by \citet{Schreiber15} as dashed lines. Galaxies above the MS are on average hotter than those on or below. Massive MS galaxies are also colder than low mass MS galaxies. {\it Bottom panels:} As in the top panels but colouring by the median dust mass, as indicated by the colour bar. Dust mass is maximal at the MS, decreasing when going above and below, at fixed stellar mass.}
\label{tempMS}
\end{center}
\end{figure*}

\begin{figure}
\begin{center}
\includegraphics[trim=0mm 0mm 5mm 11mm, clip,width=0.45\textwidth]{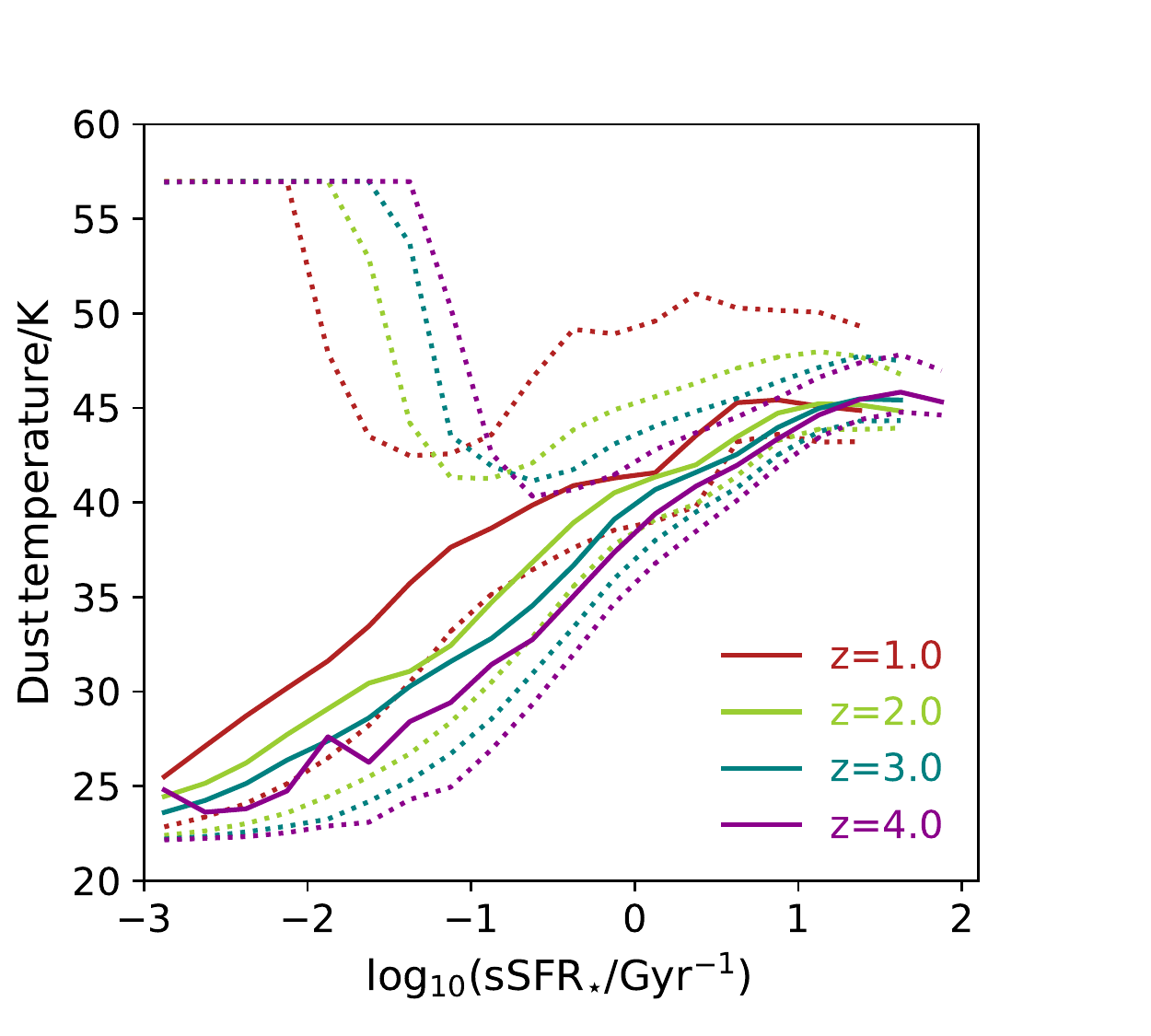}
\caption{Dust temperature vs. specific SFR at four different redshifts from $z=1$ to $z=4$, as labelled, using a $\Delta z=0.15$. Solid and dotted lines show the medians and $16^{\rm th}-84^{\rm th}$ percentile ranges in bins of sSFR, respectively. Note that at $\rm log_{10}(sSFR/Gyr^{-1})\lesssim -1$, the $84^{\rm th}$ percentile saturates at a temperature of $\approx 57$~K. This is due to galaxies having their IR emission being almost entirely dominated by birth cloud emission.}
\label{TSSFR}
\end{center}
\end{figure}

The way we compute SEDs for galaxies is thoroughly described in \citet{Lagos19}; here we provide a brief overview. 

We make use of two packages: \prospect\footnote{\href{https://github.com/asgr/ProSpect}{\url{https://github.com/asgr/ProSpect}} and for an interactive \prospect\ web tool see {\url{http://prospect.icrar.org/}}, which is recommended as an education tool.} and \viperfish\footnote{\href{https://github.com/asgr/Viperfish}{\url{https://github.com/asgr/Viperfish}}}. \prospect\ \citep{Robotham20} combines the GALEXev stellar synthesis libraries \citet{Bruzual03} (BC03 from hereafter) and/or EMILES \citep{Vazdekis16} with the two-component dust attenuation model of \citep{Charlot00} and dust re-emission using the templates of \citep{Dale14}. The latter cover up to a rest-frame wavelength of $1,000\mu$m. 
On top of \prospect\ sits \viperfish, which allows for simple use of \shark\ SFHs and ZFHs, and generation of the desired SED through target filters. For this paper we include a large range of bands: GALEX FUV and NUV, SDSS u, g, r, i, z VISTA Y, J, H, K, WISE bands 1, 2, 3, 4, Spitzer IRAC bands 1, 2, 3, 4, Herschel PACS 70$\mu$m, 100$\mu$m, 160$\mu$m, and SPIRE 250$\mu$m, $350\mu$m, $500\mu$m, JCMT 450$\mu$m and $850\mu$m, and ALMA bands 9, 8, 7, 6, 5 and 4. For all, except the ALMA bands, we use the published filter responses that come with \prospect, while for ALMA we use a top-hat filter over the frequency range of each band\footnote{{The frequency range adopted can be found here \url{https://www.eso.org/public/teles-instr/alma/receiver-bands/}.}}. This is done to thoroughly study the UV-to-IR emission of submm and mm-selected galaxies. When using a lightcone, galaxy SEDs are shifted to the observer-frame according to their redshifts.
Note that due to the wavelength limit of the IR templates of \citet{Dale14}, observer-frame galaxy emission in ALMA bands 5 and 4 is only computed for galaxies at $z\ge 0.42$ and $z\ge 0.84$, respectively. We are currently working on \prospect\ to include a module for radio continuum emission from both free-free and synchrotron processes, which will naturally allow our machinery to extend our predictions to band~3 and the Square Kilometer Array frequency coverage. 

In \viperfish, we attenuate and re-emit the light due to birth clouds first, and then attenuate and re-emit the light due to the diffuse ISM. The \citet{Charlot00} (hereafter CF00) absorption curve for stars in the diffuse ISM and birth clouds can be written as follows

\begin{eqnarray}
\tau_{\rm ISM} &=& \hat{\tau}_{\rm ISM}\, (\lambda/5500\text{\AA})^{\eta_{\rm ISM}}
\label{taus_screen},\\
\tau_{\rm BC} &=& \hat{\tau}_{\rm BC}\, (\lambda/5500\text{\AA})^{\eta_{\rm BC}},\label{tau_bc}
\end{eqnarray}

\noindent respectively, where $\hat{\tau}_{\rm ISM}$ and $\hat{\tau}_{\rm BC}$ are the optical depth at $5500$\AA\ in the diffuse ISM and birth clouds, respectively, and $\eta_{\rm ISM}$ and $\eta_{\rm BC}$ are the power-law indices that control the dependence on wavelength for the diffuse ISM and birth clouds, respectively. In \shark\ we scale the CF00 parameters above depending on local properties of galaxies. 

In the case of birth clouds, we compute the optical depth from the gas 
metallicity and a typical cloud surface density (see Eq.~$6$ in \citealt{Lagos19}). The latter is $\Sigma_{\rm gas,cl}={\rm max}[\Sigma_{\rm MW,cl},\Sigma_{\rm gas}]$, with $\Sigma_{\rm MW,cl}=85\,\rm M_{\odot}\,pc^{-2}$ \citep{Krumholz09}, $\Sigma_{\rm gas}$ being the diffuse medium gas surface density, $0.5\,\Sigma_{\rm gas}/\pi\,r^2_{\rm 50}$, and 
$r_{\rm 50}$ being the half-gas mass radius. The wavelength power-law index is fixed to $\eta_{\rm BC}=-0.7$.

For the diffuse dust we use the scaling derived from the RT analysis of {\sc EAGLE} galaxies at $0\le z\le 2$ using {\sc SKIRT} \citep{Camps15} proposed by \citet{Trayford19}. These consist of $\hat{\tau}_{\rm ISM}$ and $\eta_{\rm ISM}$ varying as a function of the dust surface density, $\Sigma_{\rm dust}$, which are redshift independent. We compute $\Sigma_{\rm dust}$ independently for the disk and the bulges of galaxies. To compute $\Sigma_{\rm dust}$, we include the effect of inclination as described in \citet{Lagos19} (Eq. 2).

Dust masses in \shark\ are calculated for each galaxy component from their gas mass and metallicity, following the best fit relation between the dust mass and the latter two galaxy properties in the local Universe of \citet{Remy-Ruyer14} (see Fig.~1 in \citealt{Lagos19}). The surface density is then computed from the half-gas mass radius of the disk or bulge and the inclination in which the galaxy is seen - it therefore depends on where the observer is located. The details of these calculations can be found in $\S$~3.1 of \citet{Lagos19}. 

For the remission of the absorbed light in the IR, we use the \citet{Dale14} templates, which are parametrised by a power-law index that controls the global dust emission depending on the local interstellar radiation field U, with a fraction ${\rm d}M_{\rm dust}$ of dust mass being heated by $U^{-\alpha_{\rm SF}}{\rm d}U$, and $U=1$ being the local interstellar radiation field of the solar neighbourhood. In \viperfish, we adopt  $\alpha_{\rm SF}=3$ for the diffuse ISM and $\alpha_{\rm SF}=1$ for the birth clouds, {motivated by the $\alpha_{\rm SF}$ values derived in \citet{Dale14} for a range of galaxies going from normal star-forming (typically $\alpha_{\rm SF}\approx 2-3$) to highly starbursting (typically $\alpha_{\rm SF}\approx 1-2$)}. 
The SED model adopted here is referred to as ``{EAGLE-$\tau$ RR14}'' in \citet{Lagos19}.

From the re-emission of the absorbed light in the IR we can compute an effective dust temperature. We use the total IR luminosity emitted by the screen and birth cloud components of the IR templates of \citet{Dale14}. 
 The values adopted here for $\alpha_{\rm SF}$ for the screen and birth cloud components roughly correspond to effective dust temperatures of $22$~K and $57$~K, respectively, which we referred to as $T_{\rm diff}$ and $T_{\rm bc}$, respectively. This way, a single temperature is computed following \citet{daCunha15}:
 
 \begin{equation}
     T_{\rm eff} = \frac{\left(L_{\rm IR, diff} \, T{\rm diff} + L_{\rm IR, bc} \, T_{\rm bc}\right)}{L_{\rm IR}},\label{temp_calc}
 \end{equation}

\noindent where $T_{\rm eff}$ is the effective dust temperature, $L_{\rm IR} = L_{\rm IR, diff} + L_{\rm IR, bc}$, and $L_{\rm IR, diff}$ and $L_{\rm IR, bc}$ are the IR luminosity produced by the diffuse and birth cloud dust components, respectively. The latter is approximately equivalent to the IR emission in the $4-1000\mu$m wavelength range. 

The most simplistic assumption we make in the build-up of SEDs of \shark\ galaxies, are the fixed $\alpha_{\rm SF}$ for birth clouds and diffuse dust, which has consequences on the range of dust temperature \shark\ galaxies can have. 
The top panels of Fig.~\ref{tempMS} show the median dust temperature of galaxies in the SFR-stellar mass plane at four different redshifts, from $z=1\pm 0.15$ to $z=4\pm 0.14$. We show the median SFR of all galaxies with a $\rm SFR>0$ and the $1\sigma$ scatter around the median for reference. We consider this to be a good proxy for the MS position in \shark, except in massive galaxies, $M_{\star}\gtrsim 10^{10.2}\,\rm M_{\odot}$, at $z\lesssim 3$, which are affected by AGN feedback. The latter becomes more significant towards low redshift. For reference, we show the observational inferences of the MS from \citet{Schreiber15}, which confirm that deviations in \shark\ take place in massive galaxies. 

Lines of constant dust temperature become shallower with increasing redshift on the MS and below. Above the MS (i.e. above the $84^{\rm th}$ percentile range), lines of constant dust temperature run parallel to the MS with a weak trend of increasing temperature with increasing {SFR at fixed stellar mass}. If we focus on the MS and below, at $z=1\pm0.15$ and $z=2\pm 0.15$, the lines of constant dust temperature being steeper than the MS lead to the trend of low-mass galaxies having on average hotter dust than massive galaxies; by $z=4\pm 0.15$ lines of constant temperature are parallel to the SFR MS even in the MS and below. The trends above largely mimic the variations in the SFR surface density and specific SFR of galaxies in the SFR-stellar mass plane. Fig.~\ref{TSSFR} shows the dependence of the dust temperature on specific SFR (sSFR). Dust temperature systematically increases with increasing sSFR, which is qualitatively similar to what \citet{daCunha08} find for MAGPHYS (see their Fig.~14). There is a mild redshift evolution, in a way that at fixed sSFR, the high-z galaxies tend to have lower $T_{\rm dust}$. The distribution of dust temperature at $\rm sSFR < 0.05\,\rm Gyr^{-1}$ is highly bimodal with the FIR emission being either fully dominated by birth cloud emission or diffuse dust. The reason for this is that the SFR in these galaxies is highly stochastic. Small episodes of SF lead to FIR emission being fully dominated by birth cloud emission. However, the mode in the low sSFR regime is dominated by diffuse dust emission only.
\citet{Schreiber18} measured $T_{\rm dust}$ of MS galaxies at $0.5<z<4$ assuming a simple Wien's law, and they find that at $z\le 1$ massive galaxies ($>10^{11}\,\rm M_{\odot}$) have lower $T_{\rm dust}$ than lower mass galaxies, in qualitative agreement with our predictions. For main sequence galaxies of $M_{\star}>10^{11}\,\rm M_{\odot}$ they found $T_{\rm dust}$ to increase from $25$~K at $z=1$ to $40$~K at $z=4$ (see also \citealt{Cowley17}). In \shark, for those galaxies $T_{\rm dust}$ increases from $\approx 30$~K at $z=1$ to $\approx 42$~K at $z=4$ in broad agreement with the observations within the systematic uncertainties.

Because the FIR emission is regulated by the dust temperature and mass, it is important to also investigate the evolution of dust masses in \shark\ galaxies. 
The bottom panels of Fig.~\ref{tempMS} show the median dust mass of galaxies in the SFR-stellar mass plane at $z\approx 1-4$. The dust mass is maximal in the MS, and decreases towards higher and lower SFRs at fixed stellar mass. This is a natural outcome of the star formation model adopted in \shark. Galaxies that undergo starbursts (those triggered by galaxy mergers and global disk instabilities) have a molecular gas depletion timescale $10$ times shorter than galaxies that undergo normal star formation in their disks. This means that for the same amount of molecular gas, the SFR is $10$ times higher in a starburst galaxy. This leads to galaxies above the MS (which are mostly in the starburst mode) having lower dust masses {(due to the lower gas content)} than MS galaxies at fixed stellar mass at all redshifts. Below the MS the trend is driven by the underlying lower molecular gas masses that lead to both lower SFRs and dust masses. At fixed distance to the MS, the dust mass increases with increasing stellar mass. 

\subsection{Carbon Monoxide emission of galaxies}\label{COmodel}

We compute the CO SLED from the $1\rightarrow 0$ to the $10\rightarrow 9$ rotational transitions, using the Photon-dominated regions (PDR) modelling introduced in \citet{Lagos12}. In short, \citet{Lagos12} used a large grid of PDR models from \citet{Bayet11}, which were run using fiducial Giant Molecular Cloud (GMC) properties. These GMC models adopted a gas density  of $\rm n_{\rm H}=10^4\,\rm cm^{-4}$ and an attenuation of $A_{\rm V}=3$~mag, and were run for $10^6$~yr. These PDR models were run for a range of interstellar radiation fields, $G_{\rm UV}$, gas metallicities, $Z_{\rm gas}$ and $X$-ray fluxes, $F_{\rm X}$. In \citet{Bayet11} the latter are modelled as cosmic ray dominated regions (CRDRs), which are found to behave very similarly to X-ray dominated regions (and hence can be used in a similar fashion). The output of an individual run (which is then the combined effect of the PDR+CRDR) is $10$ molecular hydrogen-to-CO conversion factors, from $1\rightarrow 0$ to the $10\rightarrow 9$ (see Table~$1$ in \citealt{Lagos12}). 

To compute a CO Spectra Line Energy Distribution (CO SLED) for each individual \shark\ galaxy, we first compute $G_{\rm UV}$, $Z_{\rm gas}$ and $F_{\rm X}$, and interpolate over the grid of PDR models (with all properties in $\rm log_{10}$) using the {\sc Python} tool {\sc interpolate.LinearNDInterpolator} of the {\sc SciPy} package \citep{Virtanen20}. We do this for galaxy disks and bulges separately, and sum their contributions to derive a total CO SLED. Below we describe our calculation of $G_{\rm UV}$ and $F_{\rm X}$.

The interstellar radiation field relative to the solar neighbourhood value, $G_0=1.6\times 10^{-3}\rm erg\, cm^{-2}\, s^{-1}$, is computed as
\begin{equation}
    \frac{G_{\rm UV}}{G_{0}} = \frac{\rm \Sigma_{SFR}/\Sigma^0_{SFR}}{\left(Z_{\rm gas}/Z_{\odot}\right)\, \left(\Sigma_{\rm gas}/\Sigma^0_{\rm gas}\right)},
    \label{EqGUV}
\end{equation}

\noindent where $\Sigma_{\rm SFR}$ and $\Sigma_{\rm gas}$ are the SFR and gas, respectively, surface density computed at the disk and bulge half-stellar radius. $\Sigma^0_{\rm SFR}=10^{-3} \,\rm M_{\odot}\,yr^{-1}\, kpc^{-2}$ and  $\Sigma^0_{\rm gas}=10\,\rm M_{\odot}\,pc^{-2}$ are the solar neighbourhood SFR \citep{Bonatto11} and gas mass \citep{Chang02} surface densities. The physical interpretation of the numerator of Eq.~\ref{EqGUV} is that the local UV radiation field should be proportional to the SFR surface density as young stars provide the bulk of the UV emission, while the denominator accounts for attenuation. In a slab, the transmission probability of UV photons, $\beta_{\rm UV}$, scales with the optical depth, so that $\beta_{\rm UV}\sim (1 - e^{- \tau_{\rm UV}})/\tau_{\rm UV}$. The optical depth, on the other hand, depends on the gas metallicity and column density of atoms as $\tau_{\rm UV}\propto Z_{\rm gas}\, N_{\rm H}$. In optically thick gas ($\tau_{\rm UV}\gg 1$), $\beta_{\rm UV}\sim \tau^{-1}_{\rm UV}$. 

The X-ray radiation field, $F_{\rm X}$ is computed from the hard ($2-10$~kEV) X-ray luminosity of AGNs, $L_{\rm X}$, computed as in \citet{Amarantidis19} (see their Eqs.~$1-3$) for \shark, divided by the surface of the sphere of radius equivalent to the disk or bulge half-stellar mass radius, $\rm 4\,\pi\,r^2_{50}$. This assumes that hard X-rays photons are not absorbed. 

Some important caveats in the CO SLED modelling above are worth discussing.
We assume galaxy inclination can be ignored {for simplicity, as it allows us to use pre-computed PDR models}. This is likely not correct as clouds in a disk can lead to enhanced shielding between different clouds. This may be an important effect (especially at high-z) given the high CO opacity of the main isotopes. Another important caveat is that GMCs that dominate star formation at $z\approx 2$ can be $\approx 100-1000$ times more massive than local GMCs \citep{Swinbank10}. This may lead to increased self-shielding inside GMCs. 
{Both cases (enhanced self-shielding or cloud-to-cloud shielding) could lead to the CO luminosity becoming fainter.}
Despite these caveats, we show that \shark\ can broadly reproduce the observed CO SLEDs in normal star-forming galaxies at $z=1-3$ (Hamanowicz et al. in preparation) and SMGs ($\S$~\ref{cosec}).

Fig.~\ref{tempMSCO} shows the median brightness CO(1-0) luminosity, $L^{\prime}_{\rm CO(1-0)}$ of galaxies in the SFR-stellar mass plane at $z=1-4$. $L^{\prime}_{\rm CO(1-0)}$ varies in complex ways in this plane: at fixed stellar mass  $L^{\prime}_{\rm CO(1-0)}$ increases with increasing SFR up to the MS. Above the MS, $L^{\prime}_{\rm CO(1-0)}$ decreases for a bit, followed by an increase in a way that the most starbursting galaxies have the highest $L^{\prime}_{\rm CO(1-0)}$. This complex behaviour is driven by the combination of variations in the H$_2$ mass and in the conversion factor from H$_2$ to CO(1-0). Note that $L^{\prime}_{\rm CO(1-0)}$ varies in different ways than $M_{\rm dust}$ in the SFR-stellar mass plane in \shark. 

\section{SMG number counts and redshift distributions}\label{overallstats}

\begin{figure*}
\begin{center}
\includegraphics[trim=23mm 5mm 28mm 20mm, clip,width=0.97\textwidth]{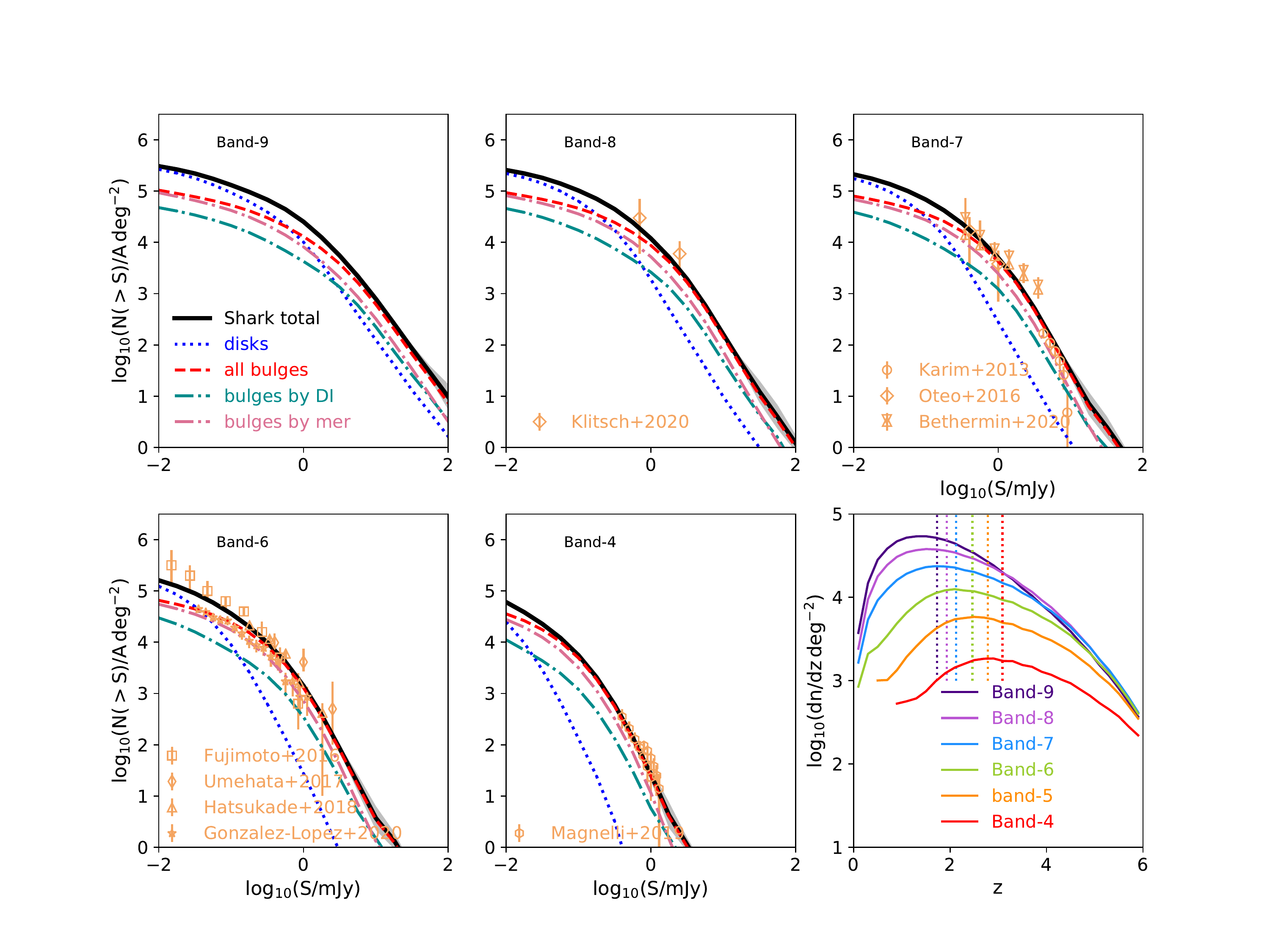}
\caption{Number counts for our \shark\ $107$~deg$^2$ deep lightcone at the ALMA bands 9, 8, 7, 6 and 4, as labelled in each panel. The contribution from disks and bulges are shown as dotted and dashed lines. We split starbursts further into those driven by galaxy mergers and disk instabilities and show them as dot dashed lines as labelled. We compute a bootstrap error on the total number counts and present those as shaded grey region. Observations from \citet{Klitsch20} for band-8, \citet{Karim13}, \citet{Oteo16} and \citet{Bethermin20} for band-7, \citet{Fujimoto16}, \citet{Umehata17}, \citet{Hatsukade18}, \citet{GonzalezLopez20} for band-6 and \citet{Magnelli19} for band-4 are shown as symbols. {For \citet{Bethermin20} we show two measurements: all their detections (down-pointing triangles) and those considered secured (up-pointing triangles).} The bottom-right panel shows the predicted redshift distribution of $>0.1$~mJy sources selected in ALMA bands 9 to 4, as labelled. {Vertical dotted lines show the respective median redshifts.} \shark\ shows that the longer the wavelength used for the selection, the higher the peak redshift of the galaxies.}
\label{numbercounts}
\end{center}
\end{figure*}

\begin{figure*}
\begin{center}
\includegraphics[trim=0mm 5mm 0mm 0mm, clip,width=0.95\textwidth]{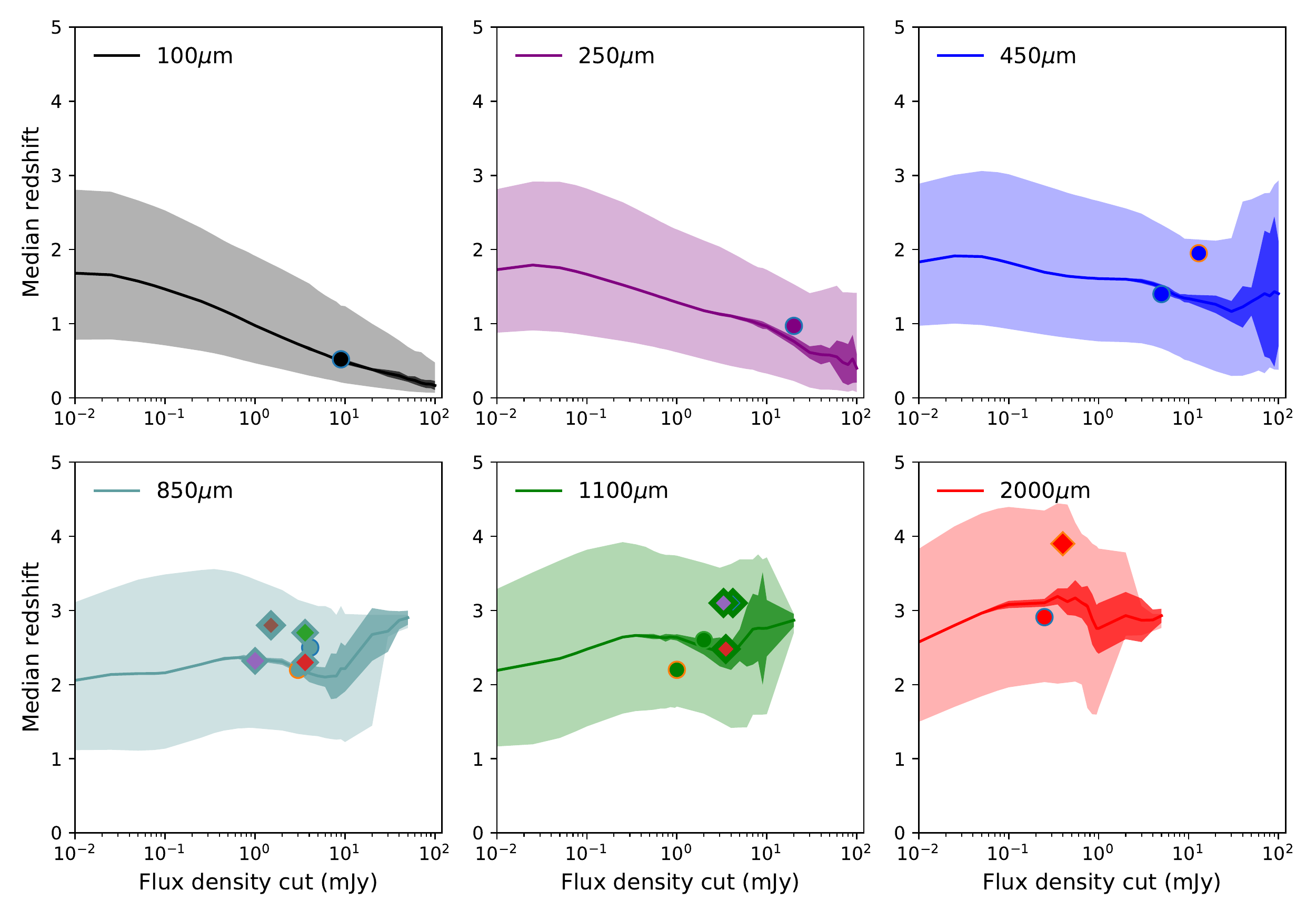}
\caption{Median redshift as a function of selection flux, for $6$ FIR bands from $100\mu m$ to $2$mm, as labelled in each panel for \shark. Dark and light shaded regions show the bootstrap error on the median and the $16^{\rm th}-84^{\rm th}$ percentile levels, respectively. Symbols in each panel correspond to observational measurements compiled by \citet{Hodge20}: $100\mu$m and $250\mu$m correspond to \citet{Berta11} and \citet{Bethermin12}, respectively, $450\mu$m correspond to \citet{Geach13} and \citet{Casey13}, 
$850\mu$m correspond to \citet{Wardlow11,Chapman05,daCunha15,Simpson14,Cowie18} and \citet{Dudzeviciute20}, $1.1$mm correspond to \citet{Smolcic12,Michalowski12,Yun12,Miettinen15,Brisbin17} and $2$mm corresponds to \citet{Staguhn14,Magnelli19}.}
\label{zdist}
\end{center}
\end{figure*}

In order to select a population of SMGs in \shark, we make use of the deep simulated lightcone introduced in \citet{Lagos19}. This corresponds to an area of $107$~deg$^2$ including all galaxies with a dummy magnitude, computed assuming a stellar mass-to-light ratio of $1$, $<32$ and at $0\le z\le 6$. \citet{Chauhan19} provide a detailed description of the construction of lightcones using \shark\ and {\sc Stingray}. The latter is the lightcone software originally developed by \citet{Obreschkow09e} and further developed by Obreschkow et al. (in preparation\footnote{\url{https://github.com/obreschkow/stingray/}}). {The exact way the simulated box at different snapshots were tiled and the angular momentum vector of the galaxies determine the inclination in which galaxies are viewed by the observer. The latter are considered in the calculation of dust attenuation ($\S$~\ref{SEDs}).}

We use this lightcone to first analyse the overall number counts and redshift distributions of galaxies selected using different IR bands and flux cuts ($\S$~\ref{overallstats}). We then focus on the UV-to-MIR properties of ALMA-selected galaxies in \shark\ ($\S$~\ref{SEDPropsSMGs}). Finally, we focus on the intrinsic properties of SMGs and their environments throughout cosmic time ($\S$~\ref{intrinsicprops}). 

\citet{Lagos19} showed that the predicted GALEX NUV to JCMT $850\mu$m number counts agreed well with observations. Here, we extend this testing of \shark\ to a large range of ALMA bands, from band 9 ($\lambda=[0.4,0.5]\,$mm) to band 4 ($\lambda=[1.8,2.4]\,$mm). ALMA is the best instrument to compare our direct predictions of galaxy number counts with as its excellent angular resolution allows a robust identification of the emission of individual galaxies \citep{Karim13}. Previous estimates from instruments such as JCMT suffered from significant blending due to their poor angular resolution, forcing simulated lightcones to account for such blending (e.g. \citealt{Cowley15}). By comparing with ALMA derived number counts we are able to bypass this issue and compare directly with our predictions.

Fig.~\ref{numbercounts} shows the number counts of FIR sources in our simulated lightcone from band~9 to band~4. Observational constraints are currently available for bands~8, 7, 6 and 4 (among the bands we can reliably predict SEDs for) and hence the rest serves as predictions. At band~4, we show the observational estimates of \citet{Magnelli19}, which come from the GIZMO instrument rather than ALMA, but are at band~4 wavelength. We find that \shark\ predicts numbers counts that are in excellent agreement with observations. There is a significant tension between different observational datasets in band~6, that is highly apparent at fluxes $<0.7$~mJy. At brighter fluxes the differences can be attributed to small number statistics and are typically within the errorbars. \shark\ predictions lie in between these different observations and within their uncertainties, although its closer to the number counts presented in \citet{Fujimoto16}. The area of our simulated lightcone is more than $4$ orders of magnitude larger than what can be accessed with ALMA. For reference, the ALMA large program ASPECS from where the number counts of \citet{GonzalezLopez20} come from has an area of $\approx 0.006$~deg$^2$, while our lightcone has an area of $107$~deg$^2$.
Hence, it is likely that the tension between these different data sets comes from cosmic variance. For reference, using the cosmic variance calculator of \citet{Driver10}\footnote{Available from \url{http://cosmocalc.icrar.org/}.} a survey of the area of ASPECS and in the redshift range of $1<z<3$ has a cosmic variance of $\approx 20$\% (however, see \citealt{Popping19} for a counter argument). {\shark\ predicts a weak upturn in the abundance of galaxies at the bright end. This is caused by bright IR galaxies  at low redshift (see Figure~17 of \citealt{Lagos19}). \citet{Lagos19} compared the predicted number counts of \shark\ with Herschel observations and there are indications that the abundance of bright galaxies may be too high in the model, perhaps indicating that \shark\ tends over-predicts the abundance of local starbursts. This is qualitatively similar to what {\sc EAGLE} predicts, but in \shark\ it manifests to a much lesser degree.}

In addition to showing the total number counts in Fig.~\ref{numbercounts}, we also show the contribution from disks and bulges. The latter refer to the components of a galaxy, and hence a single galaxy with both components would contribute to the dotted and dashed lines.
In all wavelengths studied in Fig.~\ref{numbercounts}, disks dominate at the faint-end. However, the transition from bulges dominating the number counts to disks happens at fainter fluxes as we move to longer wavelengths. By band-6, we expect the majority of detected galaxies in ALMA to be bulges. Because these bulges are highly star-forming, we refer to them as starbursts below. We break down starbursts into their driving mechanism: galaxy mergers and disk instabilities. Note that from a physical perspective we expect the latter to contain significant rotation \citep{Bournaud11} - though this is not modelled in detail in \shark. At the bright-end of the number counts, both galaxy merger- and disk instability-induced starbursts contribute in similar numbers, while mergers become more prominent at intermediate fluxes. Observations have reported that band-6 selected SMGs have mixed morphologies with some clearly displaying merger-induced features, while others being more consistent with thick disks (e.g. \citealt{Hodge19,Gullberg19}). \citet{Gullberg19} analysed the morphology of {$153$} bright {band-7} selected SMGs (average band-7 fluxes of $5$~mJy) and found the sample to have axes ratios and S\`ersic indices that are typical of bars. \citet{Seo19} in a suite of isolated Milky-Way like hydrodynamical simulations showed that bars can form quickly in gas-rich systems due to dynamical instabilities, and that these bars generally lead to the formation of bulges. If these suggested bars in observations are indeed associated to disk instabilities, then their frequency are consistent with \shark. The bottom-right panel of Fig.~\ref{numbercounts} show the redshift distributions of $>0.1$~mJy sources selected in different ALMA bands. There is a systematic shift towards higher redshift when longer wavelengths are used to select galaxies.

The agreement between \shark\ and the observations shown in  Fig.~\ref{numbercounts} is unprecedented for galaxy formation simulations, and to our knowledge, \shark\ is the first galaxy formation simulation (SAM or hydrodynamical) to reach this level of agreement. We note that our model was not tuned to reproduce galaxy number counts, and our SED construction was simply done using state-of-the-art methods and physical insight from RT calculations of hydrodynamic galaxies. Hence, the agreement of Fig.~\ref{numbercounts} was not guaranteed and is therefore a true predictive success of our model. We also refer the reader to \citet{Lagos19} for a comparison with observed number counts and luminosity functions over a wide wavelength range, from the UV to the FIR. 

Another independent test of galaxy formation simulations is the redshift distribution of FIR-selected sources. Fig.~\ref{zdist} shows the medians and $16^{\rm th}$-$84^{\rm th}$ percentiles of galaxy samples selected at wavelengths from $100\mu$m to $2$mm to be brighter than a given flux threshold. By doing this we are assuming that a survey is $100$\% complete down to that flux threshold. Symbols show the observational compilation of \citet{Hodge20}. The only difference with the latter is that we adapted the flux density cut values of \citet{Cowie18,Magnelli19} to better reflect their minimum flux detected. At wavelengths $100\mu$m and $250\mu$m, \shark\ predicts that the deeper the survey the higher the redshift of the selected galaxies - the brightest sources are therefore more likely to be at lower redshifts. This trend starts to revert at $450\mu$m due to negative k-correction, where \shark\ predicts a very mild dependence of the median redshift of a galaxy survey with its depth. At $850\mu$m, \shark\ predicts a very flat median redshift dependence on the survey depth, except at the brightest flux density thresholds, where most galaxies are expected to be at higher redshifts (see also bottom-right panel of Fig.~\ref{numbercounts}). At $1.1$mm {(band-6)} and $2$mm {(band 4)}, there is a positive dependence between the median redshift of a survey and its flux selection, so that the highest redshift galaxies are also the brighter ones. The case of $2$mm {(band 4)} selected galaxies is interesting, as \shark\ shows that a depth of $\approx 0.3$~mJy is the optimal one to maximise the chance to have a significant number of galaxies at $z>4$. A much brighter cut would result in the loss of most of the $z>4$ tail.
Compared to the observational compilation of \citet{Hodge20} we find that \shark\ agrees reasonably well - however some systematic differences are worth mentioning. First at band~7 and band~6, it appears that the mode of the predicted redshift distribution is slightly lower compared to some of the observations. Some observations, however, favour a lower redshift mode, in better agreement with our predictions. At $2$mm {(band 4)} we also see significant differences in the median redshift estimates of different studies, with \shark\ preferring the lower redshift solution. It is also worth highlighting that different methods to measure photometric redshifts yield quite different results. For example, \citet{Simpson14} and \citet{daCunha15} used the same observational sample but obtained median redshifts of $2.3$ and {$2.5$}, respectively. This shows that systematic uncertainties on these redshift distributions are large as they are mostly based on photometric constraints on redshifts rather than spectroscopic confirmation. Redshift campaign follow-ups are thus extremely valuable and necessary to better constrain the simulations.

The broad agreement between the predicted number counts and redshift distributions and observations gives us confidence that we can use \shark\ and the lightcone presented here to explore the UV-to-MIR SEDs of FIR-selected galaxies as well as intrinsic properties of SMGs.

\section{UV-to-MIR properties of SMGs in \shark{}}\label{SEDPropsSMGs}

\begin{figure*}
\begin{center}
\includegraphics[trim=20mm 15mm 27mm 22mm, clip,width=0.75\textwidth]{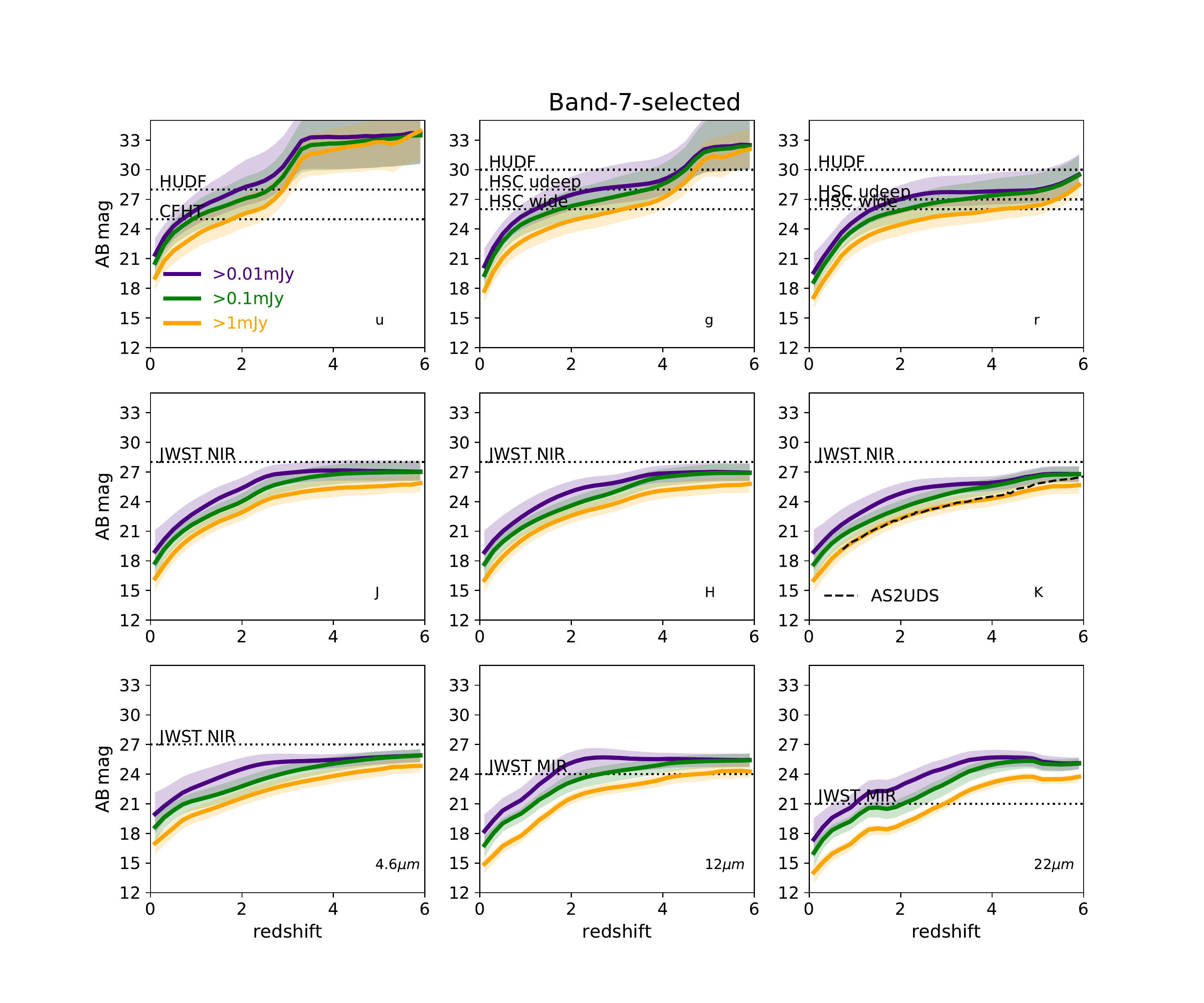}
\caption{Apparent magnitude of \shark\ SMGd in the optical (u, g, r), NIR (J, H, K, $4.6\mu$m) and MIR ($12\mu$m and $22\mu$m) bands as a function of redshift. SMGs are selected from their ALMA band-7 continuum emission, adopting $3$ different threshold flux above which galaxies are selected (labelled in the top-left panel). Lines with shaded regions show the medians and $16^{\rm th}-84^{\rm th}$ percentile ranges, respectively. We show as horizontal lines some various flux limits of relevant surveys, including HUDF (\citealt{Beckwith06}), the HSC wide and ultra-deep surveys \citep{Aihara19}, the COSMOS CFHT survey \citep{Capak07}, and fiducial $10,000$~seconds integration with the MIRI and NIR JWST instruments. For reference we also show the $K$-band vs. redshift track from the composite SED of the AS2UDS band-7 SMG sample of \citet{Dudzeviciute20}, which have a median flux of $\approx 2-4$~mJy, and hence should be comparable to the \shark\ sample with $S_{\rm band-7}>1$~mJy.}
\label{ncountsband7selec}
\end{center}
\end{figure*}

An important area of research in SMGs has been the measurement of their panchromatic emission. In order to explore the optical-to-MIR emission of \shark's SMGs, we analyse the $u$, $g$, $r$, $J$, $H$, $K$, $4.5\mu$m, $12\mu$m and $22\mu$m  apparent magnitude for band-7 selected sources in Fig.~\ref{ncountsband7selec}. 
We show as horizontal lines the estimated sensitivity for several existing and future surveys. The magnitude limits of the Hubble Ultra Deep Field (HUDF; \citealt{Beckwith06}) are shown in the top three panels; the COSMOS Canada-French Hawaii Telescope (CFHT) survey \citep{Capak07} is shown in the top-left panel; the Hyper Suprime-Cam (HSC) Subaru wide and ultra-deep surveys are shown in the top-middle and top-right panels \citep{Aihara19}; a fiducial James Webb Space Telescope (JWST) survey with the near-infrared camera (NIR) and mid-infrared instrument (MIRI), integrating for $10,000$ seconds with each and taking as a threshold a $\rm S/N = 10$. The latter depth were computed for a point source. Several ALMA large programs have observed in the HUDF and COSMOS regions, and hence these survey limits are of particular interest in this analysis. We find that HUDF should provide $r$-band counterparts for all SMGs, with the exception of some of the highest redshift, $z>5$, sources with fluxes $>0.01$~mJy. In the $u$-bands and $g$-bands, however, HUDF would miss all sources with $S_{\rm band-7}>1$~mJy at $z\gtrsim 3$ and $z\gtrsim 5$, respectively. Going deeper in band-7 flux only makes this worse: sources with  $S_{\rm band-7}>0.01$~mJy would have no HUDF $u$ and $g$ counterparts at $z\gtrsim 2$ and $z\gtrsim 4$, respectively. 

The HUDF survey comprises a very small area of $0.00127$deg$^2$ and hence is not ideal for the study of SMGs (given the small numbers per unit area expected; see Fig.~\ref{numbercounts}). Larger area surveys, such as HSC's wide ($1,400$deg$^2$), deep ($27$deg$^2$) and ultra-deep ($3.5$deg$^2$) surveys likely offer the current best compromise of depth and area for the study of high-redshift sources. The top panels of 
Fig.~\ref{ncountsband7selec} show the magnitude limits of HSC's wide and ultra-deep surveys. In the $r$-band, these surveys should provide counterparts for sources with  $S_{\rm band-7}>1$~mJy at $z\lesssim 4$ and $z\lesssim 6$, respectively; while at the $g$-band, counterparts should be detected for $S_{\rm band-7}>1$~mJy sources at $z\lesssim 3$ and $z\lesssim 4.5$, respectively. Focusing on fainter band-7 sources, $S>0.01$~mJy, we find that HSC's wide (ultra-deep) would provide $g$ and $r$-band counterparts for sources at $z\lesssim 1$ ($z\lesssim 2$) and $z\lesssim 1.7$ ($z\lesssim 4$), respectively. In the NIR, a fiducial $10,000$s JWST NIR survey would be capable of comfortably detecting counterparts for all band-7 selected sources, even as deep as $S_{\rm band-7}>0.01$~mJy, in the $J$, $H$ and $K$ bands, and even up to $z=6$. Similar integration times with JWST's MIRI instrument though would provide counterparts for $S_{\rm band-7}>1$~mJy sources only at $z\lesssim 4$ at $12\mu$m and $z\lesssim 3$ at $22\mu$m. For fainter band-7 sources this becomes worse, typically only detecting counterparts for $S_{\rm band-7}>0.01$~mJy sources at $z\lesssim 1.5$.

To investigate the consistency of our predicted magnitudes with observations, we turn our attention to the optical-NIR analysis of the ALMA survey AS2UDS sources presented in \citet{Dudzeviciute20}. \citet{Dudzeviciute20} analysed sources with band-7 fluxes $>0.6$~mJy, but with the mode being at $\approx 2-4$~mJy. Below we focus on the IRAC $3.6\mu$m band as \citet{Dudzeviciute20} showed this to have the highest completeness. The authors found that $\approx 90$\% of their sources had IRAC $3.6\mu$m counterparts brighter than $23.5$mag. In order to compare with the latter, we select galaxies in our \shark\ lightcone to have the same band-7 flux distribution and find that $89.9\pm 1.3$\% (error computed from random re-sampling) of galaxies have IRAC $3.6\mu$m emission $<23.5$~mag, in excellent agreement with the observations. Table~\ref{Iracgals} shows the median properties of band-7 galaxies with fluxes $>1$~mJy that would be IRAC-bright and faint, according to the magnitude threshold above. IRAC-faint SMGs are at significantly higher redshift, have smaller stellar and dust masses (by $\approx 0.5$~dex), higher sSFR (by a factor of $2$), and higher rest-frame $V$-band attenuation (by $0.7$~mag) than the IRAC-bright sources. Of all these properties, the ones that seem to be more fundamental in making them IRAC faint are the redshift and the rest-frame $V$-band attenuation. In $\S$~\ref{intrinsicprops} we discuss more in general how these properties evolve for galaxies selected in different ALMA bands.

\begin{table}
    \centering
    \caption{Median stellar mass ($M_{\star}$), sSFR, rest-frame V-band attenuation, $A_{\rm V}$, dust temperature ($T_{\rm dust}$), dust mass ($M_{\rm dust}$) and redshift for \shark\ galaxies that in band-7 have a flux $>1$~mJy and have IRAC $3.6\mu$m apparent magnitudes $<23.5$ (bright) and $>23.5$ (faint). The errors associated to the medians represent the $16^{\rm th}$ and $84^{\rm th}$ percentile ranges.}
    \begin{tabular}{c|c|c}
         \hline
          & $S_{\rm band-7}>1$mJy& \\
         \hline
         Property & IRAC $\rm 3.6\mu m<23.5$& IRAC $\rm 3.6\mu m>23.5$\\
         \hline
         $\rm log_{10}(M_{\star}/M_{\odot})$& $10.44^{+0.42}_{-0.39}$ & $10.05^{+0.3}_{-0.34}$\\
         $\rm log_{10}(sSFR/Gyr^{-1})$ & $0.37^{+0.46}_{-0.66}$& $0.77^{+0.34}_{-0.4}$\\
         $A_{\rm V}\rm /mag$ & $1.5^{+0.55}_{-0.5}$& $2.2^{+0.55}_{-0.46}$\\
         $T_{\rm dust}/\rm K$ & $43^{+3}_{-5}$& $43^{+1.5}_{-2.5}$\\
         $\rm log_{10}(M_{\rm dust}/M_{\odot})$ & $7.9^{+0.9}_{-0.75}$ & $7.5^{+0.65}_{-0.57}$\\
         redshift & $2.2^{+0.82}_{-0.8}$& $3.5^{+0.87}_{-0.65}$\\
         \hline
    \end{tabular}
    \label{Iracgals}
\end{table}

We also show in the middle-right panel of Fig.~\ref{ncountsband7selec} the expected variation with redshift of the $K$-band magnitude for the composite SED of the AS2UDS sources as presented in \citet{Dudzeviciute20}. We find that the latter follows very well our prediction for a band-7 sample with $S_{\rm band-7}>1$~mJy, which is the sample of Fig.~\ref{ncountsband7selec} that is most closely comparable to the 
AS2UDS sample. Both these tests show that the SMG galaxy population in \shark\ resembles the observations quite closely. 
Comparisons at other bands (e.g. $U$ or $V$) are difficult to reliably carry out due to the high incompleteness of the UDS survey down to the magnitudes investigated in \citet{Dudzeviciute20}.

Band-6 and band-4 selected sources are shown in 
Figs.~\ref{ncountsband6selec} and \ref{ncountsband4selec}, respectively. In general, at fixed flux and redshift, sources selected in longer-wavelengths have brighter optical, NIR and MIR emission. Hence, for band-6 or band-4 selected galaxies, we generally find that existing and upcoming surveys would be able to detect counterparts up to higher redshifts than those found for band-7 selected galaxies.

\begin{figure}
\begin{center}
\includegraphics[trim=8mm 0.5mm 5mm 7mm, clip,width=0.5\textwidth]{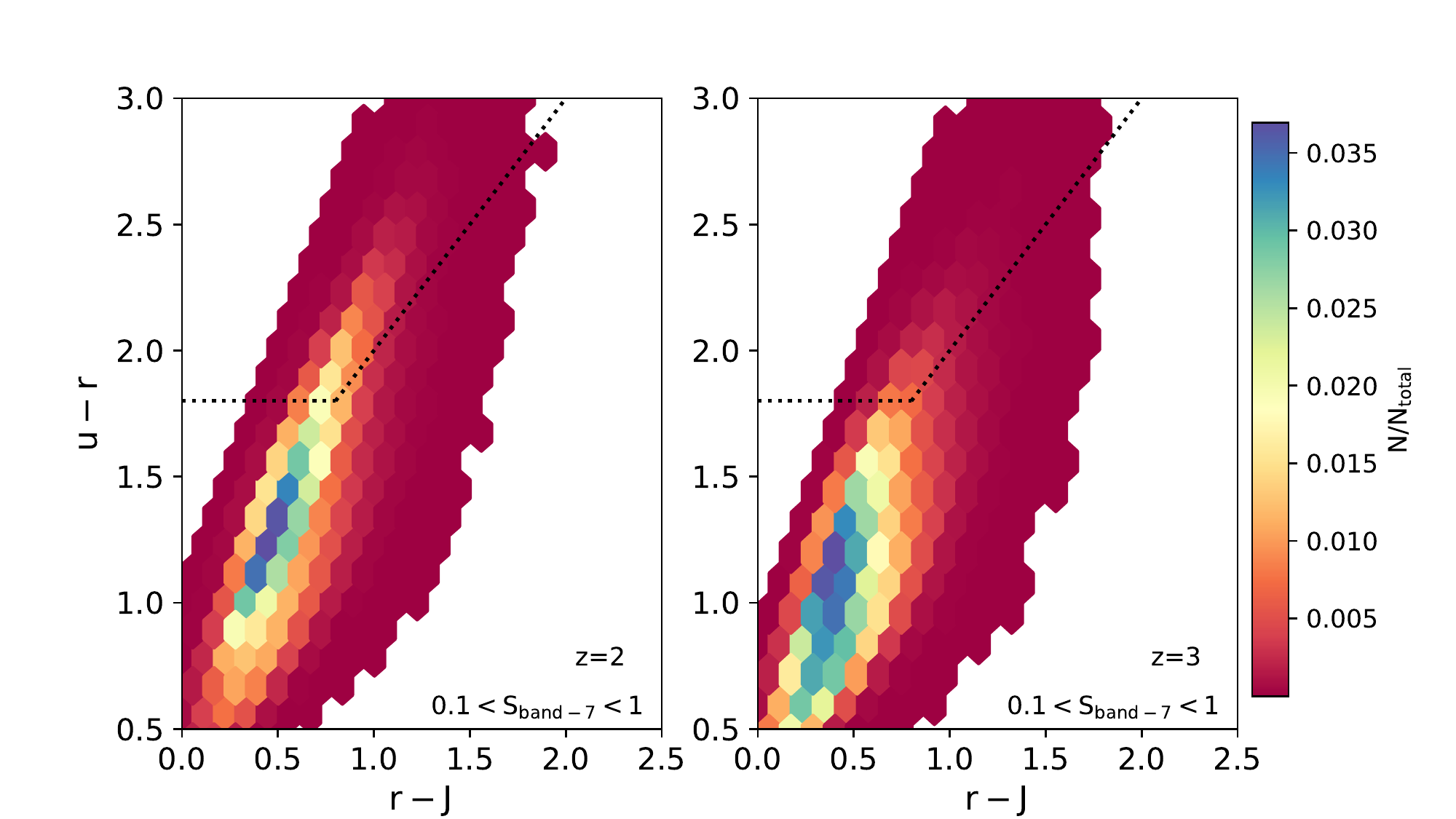}
\includegraphics[trim=8mm 0.5mm 5mm 7mm, clip,width=0.5\textwidth]{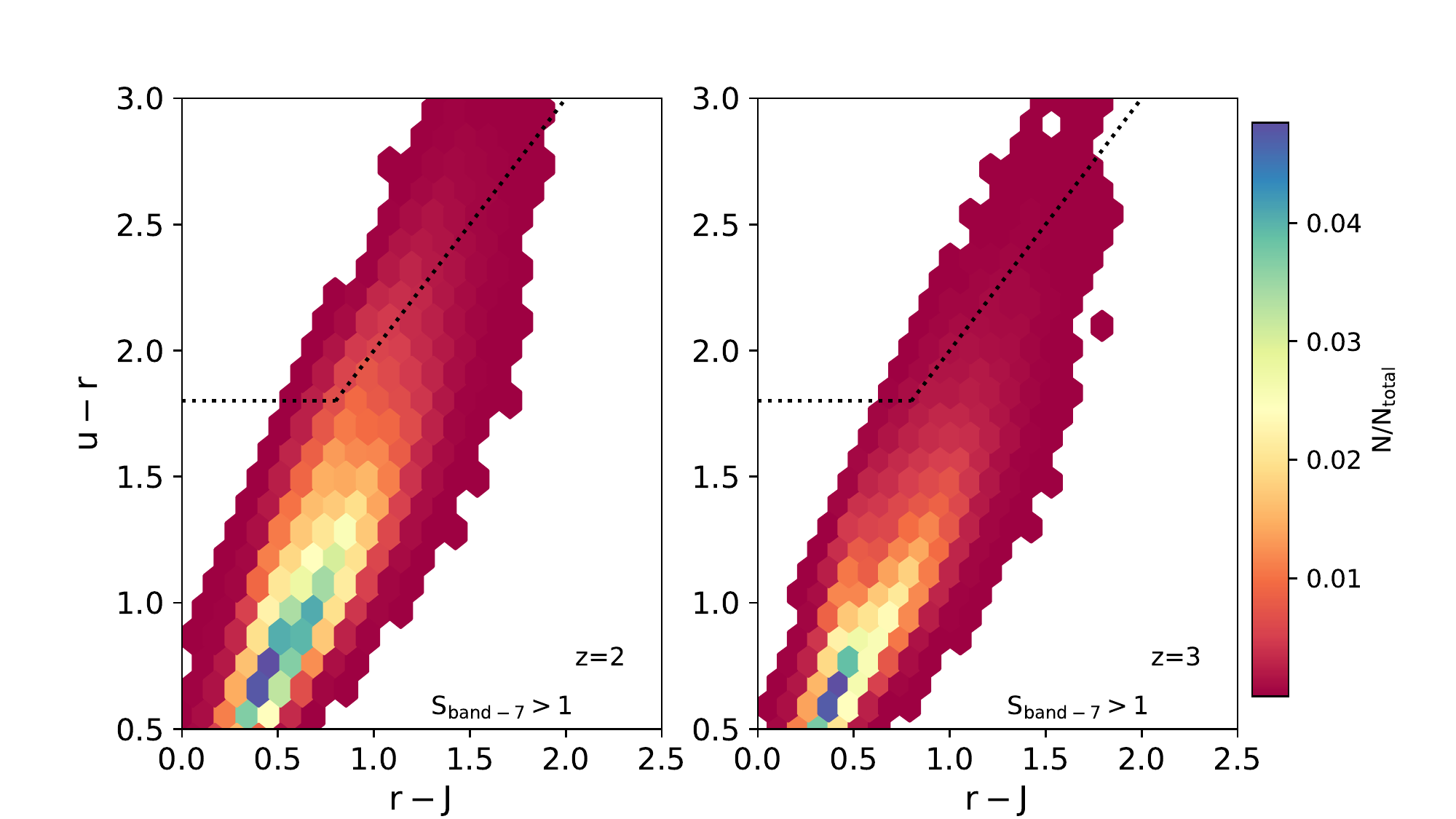}
\caption{Distribution of band 7 selected galaxies in the rest-frame (u-r) vs. (r-J) plane at redshifts $2\pm 0.5$ (left) and $3\pm 0.5$ (right). The top panel shows galaxies selected to have a band-7 flux of $[0.1,1]$~mJy, while the bottom panel showed galaxies with fluxes $>1$~mJy. The dotted lines show the area that is considered to be populated by passive galaxies based on the analysis of \citet{Bravo20} of GAMA galaxies at $z<0.5$.}
\label{smgcolours}
\end{center}
\end{figure}

\begin{figure}
\begin{center}
\includegraphics[trim=0mm 13mm 5mm 18mm, clip,width=0.45\textwidth]{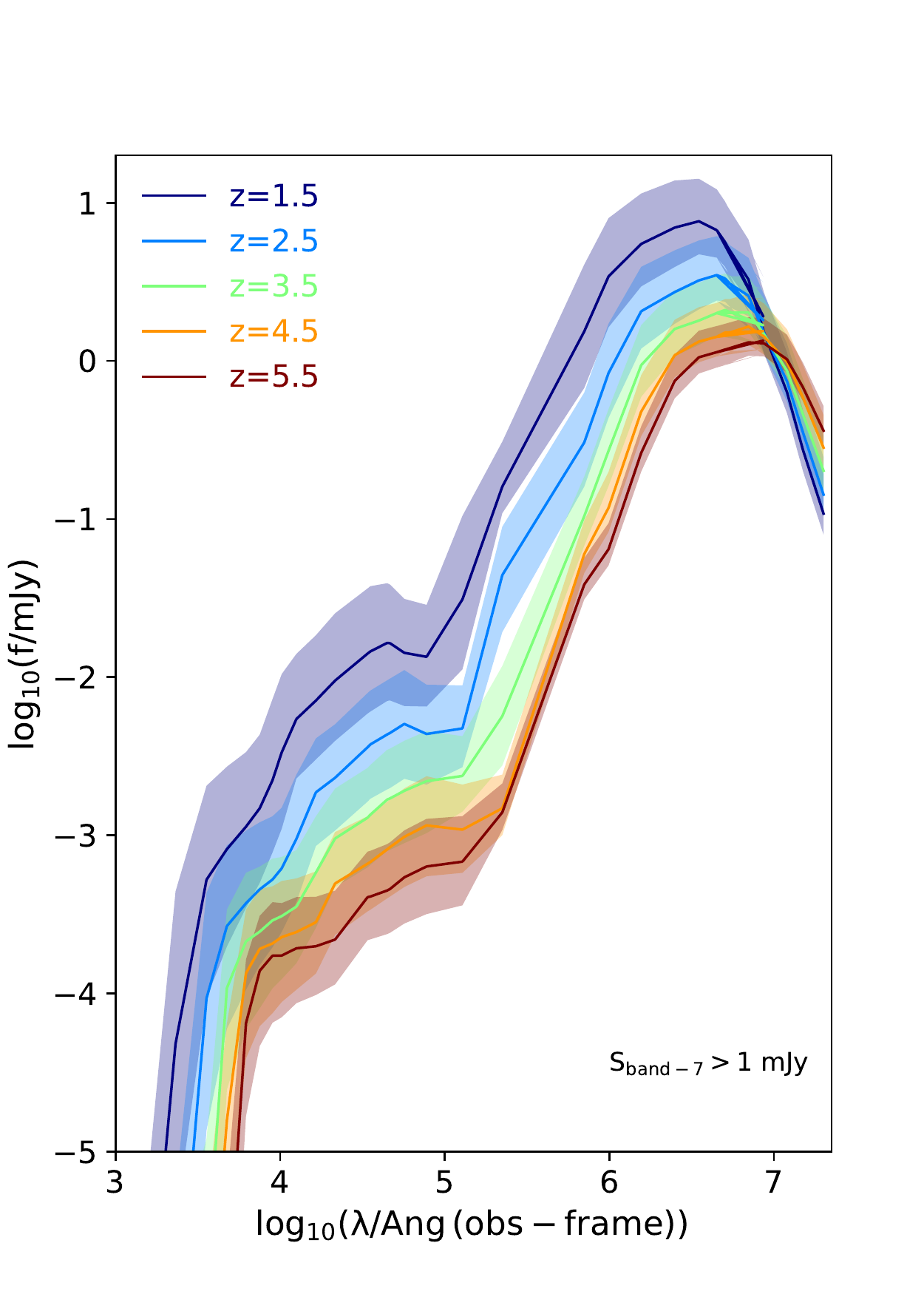}
\includegraphics[trim=-1mm 14mm 6mm 8mm, clip,width=0.45\textwidth]{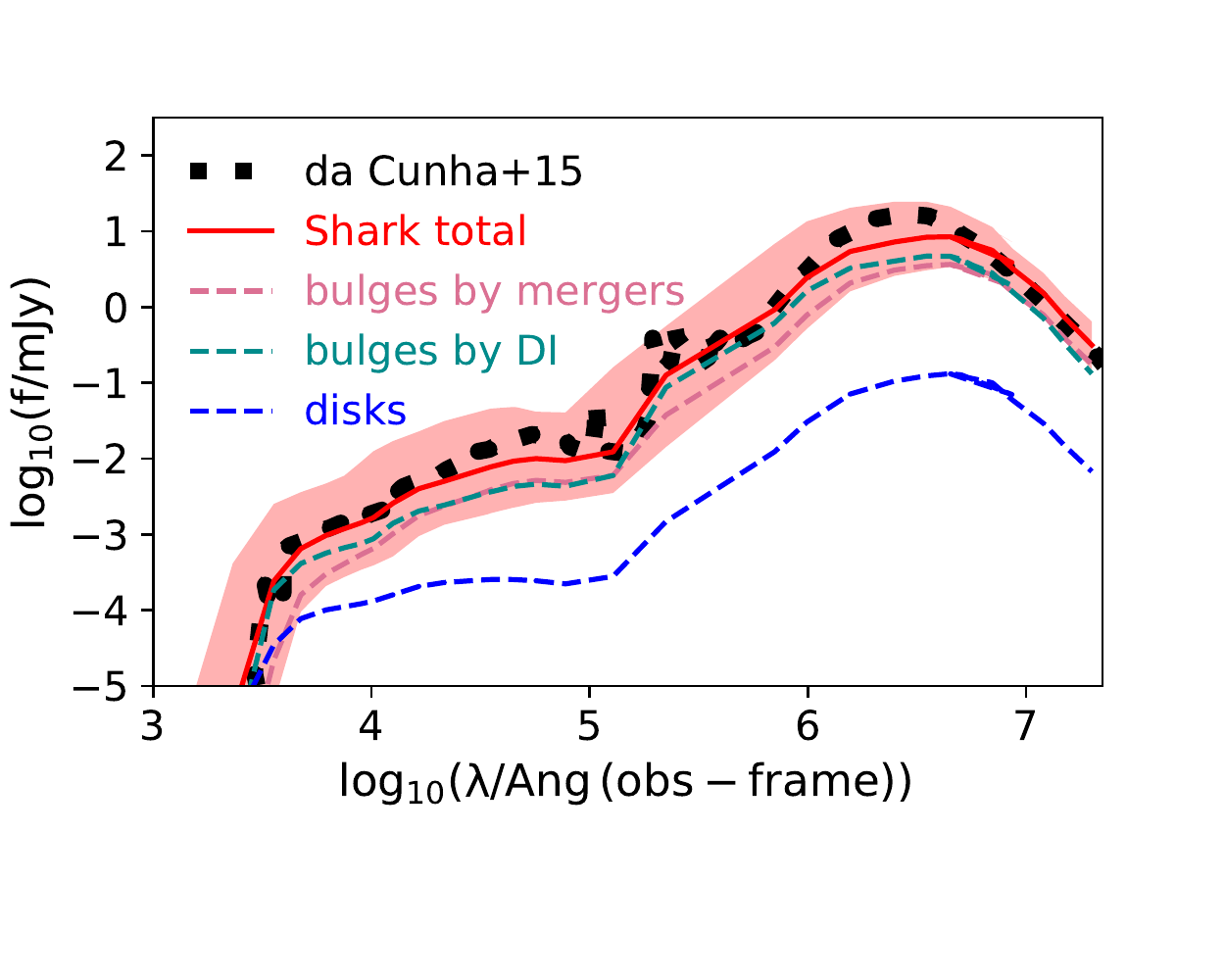}
\caption{{\it Top panel:} Median SEDs of \shark\ galaxies selected to have $S_{\rm band-7}>1$~mJy at 5 different redshift bins: $1.5\pm 0.5$, $2.5\pm 0.5$, $3.5\pm 0.5$, $4.5\pm 0.5$, $5.5\pm 0.5$, as labelled. Lines and shaded regions show the medians and $16^{\rm th}-84^{\rm th}$ percentile ranges. Higher redshift SMGs show on average more absorption in the optical, though significant scatter in seen in the SEDs at the optical-NIR range. {\it Bottom panel:} Comparison of the observed ALESS median SED of \citet{daCunha15} with an ALESS-like sample of \shark. The latter is simply selected to have the same flux distribution as the observations. The solid line shows the median, while the shaded region shows the $16^{\rm th}-84^{\rm th}$ percentile range of the \shark\ prediction. The dashed lines show the median contribution to the total SED from bulges built by galaxy mergers, disk instabilities (both classified as burst mode) and disks, as labelled.}
\label{smgseds}
\end{center}
\end{figure}

Because SMGs are highly dust obscured galaxies, their optical colours are expected to be red, potentially contaminating the typical colour-colour selections applied in observations for the selection of passive galaxies at $1\lesssim  z\lesssim 3$ (e.g. \citealt{Daddi04,Brammer11,Leja19}) and even at $z\gtrsim 5$ \citep{Matawari20}. This is a related question to that addressed above: what is the optical-to-NIR emission of SMGs in \shark? We address this by studying the rest-frame $\rm (u-r)$ vs. $\rm (r-J)$ colour-colour plane. \citet{Bravo20} showed that this plane was very efficient at separating passive from star-forming galaxies in the GAMA survey \citep{Driver09}, and the boundaries used resemble those adopted in high redshift studies \citep{Brammer11}. Fig.~\ref{smgcolours} shows the 2D histogram in the above colour-colour plot at $z=2\pm 0.5$ and $z=3\pm 0.5$ for galaxies in \shark\ selected to have band-7 fluxes in the range $0.1-1$~mJy (top panels) and $>1$~mJy (bottom panels). We find that $\approx 4$\% and $1$\% of \shark\ galaxies with $0.1\,\rm mJy<S_{\rm band-7}<1\,\rm mJy$ would be wrongly classified as passive from their rest-frame $\rm (u-r)$ vs. $\rm (r-J)$ colour-colour position at $z=2\pm 0.5$ and $z=3\pm 0.5$, respectively. If we instead look at {\it all} galaxies that fall in the passive region of the $\rm (u-r)$ vs. $\rm (r-J)$ colour-colour plane, we find that $42$\% and $47$\% have $0.1\,\rm mJy<S_{\rm band-7}<1\,\rm mJy$ {at $z=2\pm 0.5$ and $z=3\pm 0.5$, respectively}. This percentage is smaller at $z=1$, $23$\%, but increases to $50$\% at $z=4-5$.

For the sample of \shark\ galaxies with $S_{\rm band-7}>1\,\rm mJy$, we find that only $1.6$\% and $0.4$\% of them would be wrongly classified as passive from their $\rm (u-r)$ vs. $\rm (r-J)$ colour-colour position at $z=2\pm 0.5$ and $z=3\pm 0.5$, respectively. This is much lower than the percentage found for \shark\ galaxies with $0.1\,\rm mJy<S_{\rm band-7}<1\,\rm mJy$. Of all the galaxies classified as passive, only $1.2$\% and $1.5$\% have $S_{\rm band-7}>1\,\rm mJy$ at $z=2\pm 0.5$ and $z=3\pm 0.5$, respectively. This percentage decreases mildly towards higher redshift, reaching $0.8$\% at $z=5-6$.
{Appendix~\ref{AppColsSMGs} shows the median intrinsic properties of \shark\ $S_{\rm band-7}>1\,\rm mJy$ galaxies that fall inside/outside the passive region in Fig.~\ref{smgcolours}. On average the subset that falls in the passive region has higher stellar mass and dust-to-stellar mass ratio, but lower sSFR (though still in the main sequence), compared to those that are outside this region, at fixed redshift.}

Put together {and considering all galaxies with $S_{\rm band-7}>0.1\,\rm mJy$, the fractions presented above} mean that the passive classification in a single colour-colour plane becomes more contaminated as the redshift increases {on average}, and additional photometric information is therefore required to disentangle truly passive from dust-obscured galaxies (see \citealt{Leja19} for a discussion of alternatives). However, most of the contamination is from mildly dust-obscured galaxies ($A_{\rm V}\approx 0.7-1.2$) rather than highly dust-obscured ones ($A_{\rm V}\approx 1.5-2.1$). 

The final investigation we do in this section is the evolution of the average SEDs of galaxies selected to have $S_{\rm band-7}>1\,\rm mJy$, from $z=1$ to $z=6$. The top panel of Fig.~\ref{smgseds} shows the median and the $1\sigma$ percentile range of the observer-frame FUV-to-FIR SEDs of \shark\ galaxies with $S_{\rm band-7}>1\,\rm mJy$ in $5$ bins of redshift, $z=1.5\pm 0.5$, $z=2.5\pm 0.5$, $z=1.5\pm 3.5$, $z=1.5\pm 4.5$, $z=1.5\pm 5.5$. We find that the FUV-to-NIR emission relative to the FIR emission is fainter for higher redshift SMGs, a direct result of the higher average attenuation of the high-redshift SMGs compared to their low redshift counterparts at fixed flux (see middle-left hand panel in Fig.~\ref{propssmgs2}; we come back to this in $\S$~\ref{intrinsicprops}). We also find that as redshift increases there is less variation in the FIR SED of band-7 SMGs (as seen by the smaller $1\sigma$ percentile range), but significant variations continue to be seen in the FUV-to-NIR. This then explains why band-7 SMGs span such a wide area in the $\rm (u-r)$ vs. $\rm (r-J)$ colour-colour plane of Fig.~\ref{smgcolours}. 

To compare with the median SED of observed SMGs, we take the band-7 ALESS sample of \citet{daCunha15}, and select galaxies in \shark\ to have the same band-7 flux distribution as the observed galaxies. No other constraint is applied. Hereafter, we refer to this sample as the  ``ALESS-like \shark\ sample''. We then compute the median SED of those and the $1\sigma$ scatter. The bottom panel of Fig.~\ref{smgseds} shows our prediction and the observed median SED of galaxies. \citet{daCunha15} presented their median SED as a function of rest-frame $\lambda$, which we convert to an observer-frame $\lambda$ by taking the median redshift of the sample, $z=2.2$. The agreement between the observed median SED and our predictions is remarkable. The only difference worth mentioning is the fact that the observed median SED has a slightly brighter FIR peak and at a slightly shorter wavelength than our predicted median, though comfortably within the scatter. We remind the reader though that part of the galaxy SEDs in ALESS does not come from direct measurements, but instead from the best fitting SED. Generally only upper limits are available around the peak of the FIR SEDs of ALESS galaxies.

The bottom panel of Fig.~\ref{smgseds} also shows the median contribution from bulges built via disk instabilities and galaxy mergers (both burst mode) and disks. The emission from disks plays a minor role in the median SED of ALESS-like galaxies, except at the FUV. Bulges being built via disk instabilities and galaxy mergers have similar contributions in the NIR and FIR, at wavelengths above the peak, while the former dominates in the NUV-optical and MIR-FIR up to the IR SED peak. This reinforces the idea that the bright band-7 SMGs are an even mix between starbursts driven by galaxy mergers and disk instabilities. At first glance the reader may see a contradiction between the observed morphologies of SMGs \citep{Hodge19,Gullberg19} and the fact that disks in these ALESS-like galaxies are faint. This is not the case and is simply a reflection of the simplicity with which we treat disk instabilities. In \shark, once a galaxy's  disk is found to be globally unstable, we collapse it instantaneously and drive a starburst with the available gas {that typically takes a few $100$~Myr to exhaust}. In detailed hydrodynamical simulations \citep{Bournaud10,Seo19} however, this process {of gas inflows and bulge formation} is not instantaneous, and may take several hundred Myr {(similar timescale to the molecular gas depletion)} - hence the contribution from disk instability-driven bulges should be seen as the combined contribution of the disks+bulges of those galaxies.

Overall we find that \shark\ is capable of reproducing remarkably well current constraints on the optical-to-FIR emission of SMGs. This adds weight to our predictions for what upcoming surveys, for example those performed with JWST, will be able to offer.

\section{Intrinsic properties of SMGs in \shark{} across cosmic time}\label{intrinsicprops}

In this section we focus on intrinsic properties of galaxies selected in ALMA bands 7, 6 and 4, as a function of redshift, including their environment (as measured by their host halo mass). 

\begin{figure*}
\begin{center}
\includegraphics[trim=18mm 15mm 27mm 27mm, clip,width=0.77\textwidth]{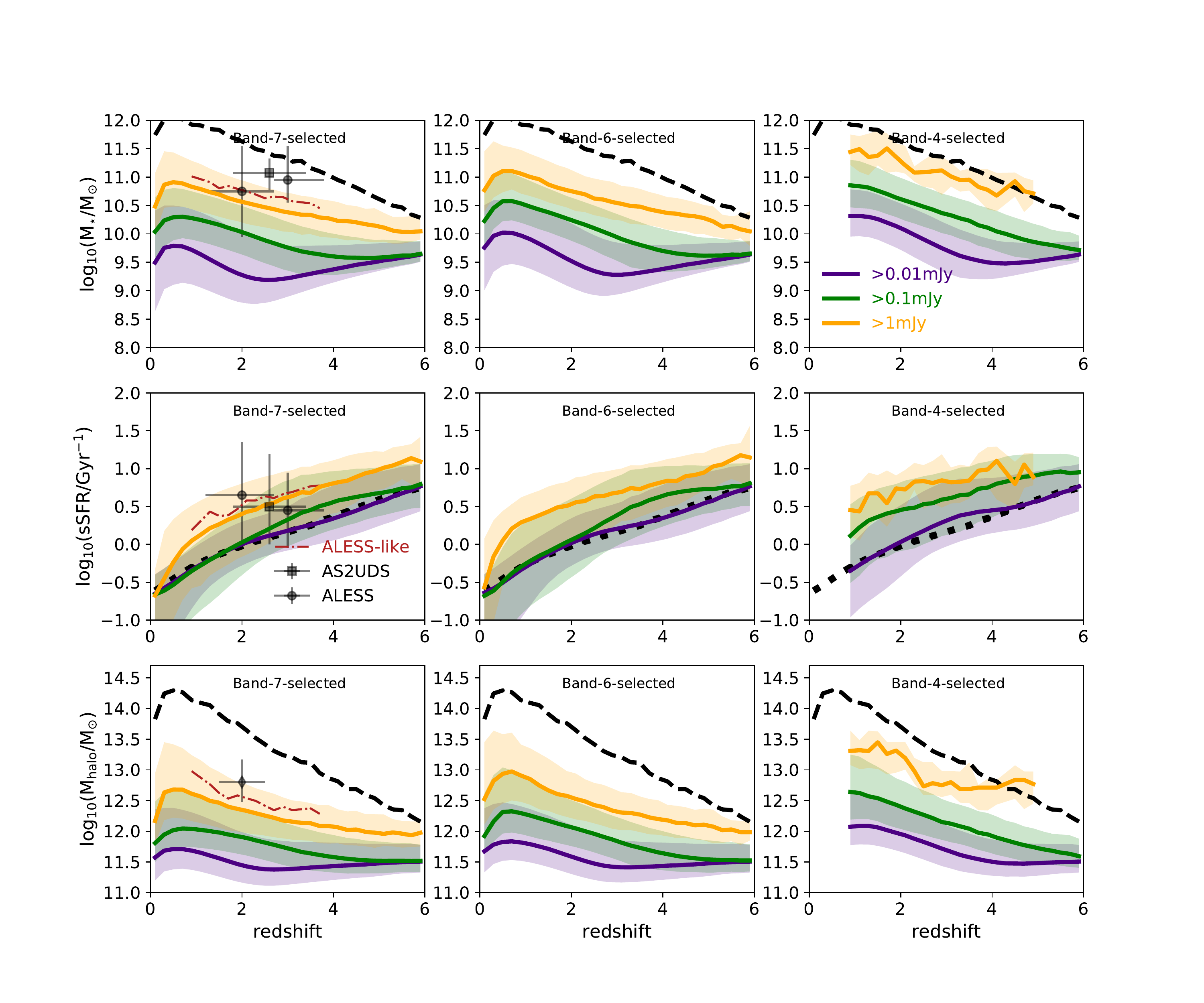}
\caption{Stellar masses, sSFRs and halo masses for the SMGs of Figs.~\ref{ncountsband7selec}, \ref{ncountsband6selec} and \ref{ncountsband4selec} as a function of redshift. Lines with shaded regions show the medians and $16^{\rm th}-84^{\rm th}$ percentile ranges, respectively. 
 For reference, the dashed lines in the top and bottom panels show the median mass of the $100$ most massive objects (in stellar and halo mass, respectively) in the simulated lightcone; the dotted lines in the middle panels show the sSFR evolution of galaxies with $10^9\,\rm M_{\odot}<M_{\star}<10^{10}\,\rm M_{\odot}$ in \shark\ that have $\rm SFR > 0$; and the dot-dashed lines in the left hand panels show the median for the \shark\ ALESS-like sample of the bottom panel of Fig.~\ref{smgseds}.
We also show the average properties of the AS2UDS \citep{Dudzeviciute20} and of the ALESS \citep{daCunha15} band-7 SMGs in the top and middle-left panels, as labelled. For ALESS, we show the two redshift bins presented in \citet{daCunha15}. At the bottom panel we show the hot halo mass inference of \citet{Hickox12} (diamond) from the spatial clustering of SMGs. All these datasets are biased towards bright band-7 SMGs, $\gtrsim 2$~mJy, and hence are more closely comparable to the \shark\ sources with $S_{\rm band-7}>1$~mJy.}
\label{propssmgs}
\end{center}
\end{figure*}

Fig.~\ref{propssmgs} shows the median stellar mass, sSFR and halo mass as a function of redshift for galaxies selected in ALMA bands 7, 6 and 4 and using three flux thresholds of $>0.01$~mJy, $>0.1$~mJy and $>1$~mJy. We also show the $1\sigma$ scatter around the medians as shaded regions. For the band-7 selected galaxies we also plot the stellar mass and sSFR of the AS2UDS sample of \citet{Dudzeviciute20} and the ALESS sample of \citet{daCunha15}, and the SMGs host halo mass inferred from clustering by \citet{Hickox12}. For band 7, we show the medians for our \shark\ ALESS-like sample (see bottom panel of Fig.~\ref{smgseds}). 
We also show for reference the sSFR evolution of MS galaxies in \shark. The latter is computed as the median sSFR of central galaxies with stellar masses $10^{9}-10^{10}\,\rm M_{\odot}$, which are vastly dominated by star-forming galaxies in \shark.
Fig.~\ref{propssmgs2} shows the median dust mass, effective dust temperature and rest-frame attenuation $A_{\rm V}$ as a function of redshift for the same \shark\ galaxies of Fig.~\ref{propssmgs}. We also show the inferred properties of ALESS galaxies presented in \citet{daCunha15} using MAGPHYS \citep{daCunha08} in the left-hand panels. 
Fig.~\ref{propssmgs3} shows the median fractional contribution of starbursts to the total SFR of galaxies, the SFR-weighted effective radii and the CO(1-0) brightness luminosity as a function of redshift for the same galaxies of Fig.~\ref{propssmgs}.
Below we analyse the relevant trends and predictions of these figures property by property.

\subsection{Stellar masses of SMGs}

Focusing first on stellar mass (top panels of Fig.~\ref{propssmgs}), we find the stellar mass to increase with increasing band-7, 6 and 4 flux. At fixed flux, band-4 galaxies are more massive than band-7 galaxies. In fact, band-4 galaxies with $S_{\rm band-4}>1$~mJy trace the most massive galaxies in the simulation at all redshits. This is seen when comparing the median for galaxies with $S_{\rm band-4}>1$~mJy with the median of the $100$ most massive galaxies in the lightcone as a function of redshift (dashed line in the top panels). Note that the latter peaks at $z\approx 1$ (rather than $z=0$) due to the limited volume probed by the lightcone at lower redshifts compared to the total volume of the simulation. This is in part responsible for the peak stellar mass we see in all the presented selections in bands 7 and 6; albeit the latter happens at slightly lower redshift, $z \approx 0.7$. The decrease in median stellar mass at $z>0.7$ in all cases is slower than the decrease in the stellar mass of the $100$ most massive galaxies, which is a consequence of ALMA bands 7, 6, 4-selected galaxies tracing more massive galaxies relative to the break of the stellar mass function as the redshift increases. At $z>4$ {\it all massive galaxies are mm bright}, which is clear from the $S_{\rm band-4}>1$~mJy sample overlapping with the dashed line in the top-right panel. 

Compared to observations, both AS2UDS \citep{Dudzeviciute20} and ALESS \citep{daCunha15} are consistent with the upper $84^{\rm th}$ percentile of the sample with $S_{\rm band-7}>1$~mJy. This is not necessarily surprising, as both samples have median band-7 fluxes of $\approx 2-4$~mJy. The median of the \shark\ ALESS-like sample is indeed higher than the $S_{\rm band-7}>1$~mJy sample by $\approx 0.2$~dex in better agreement with the observational estimates.
A large source of uncertainty in the measured properties of SMGs shown here is the lack of a secured redshift, which is a large systematic effect in stellar masses. If the redshifts were slightly over-estimated, the stellar masses of the observed SMGs would be lower and closer to the \shark\ predictions.

\subsection{The Specific Star Formation Rates of SMGs}

Focusing on sSFRs (middle panels of Fig.~\ref{propssmgs}), we find that the brighter the galaxy in ALMA bands 7, 6, 4, the higher the sSFR, with galaxies selected at longer wavelengths having higher sSFRs at fixed flux. Quantifying this in terms of the distance to the MS, we find that bands 7, 6, and 4 selected galaxies with $S>1$~mJy have median $\rm sSFR/sSFR_{\rm MS}\approx 2$, $\rm sSFR/sSFR_{\rm MS}\approx 3$ and $\rm sSFR/sSFR_{\rm MS}\approx 7$, respectively. This means that galaxies with $S_{\rm band-7}>1$~mJy and $S_{\rm band-6}>1$~mJy are only mildly above the MS, and applying some typical selection criteria in the literature they may even be considered MS - for example  \citet{Bethermin14} adopt $\rm 0.25<sSFR/sSFR_{\rm MS}<4$ to define galaxies as MS. Only band-4 galaxies with $S_{\rm band-4}>1$~mJy would truly be considered starbursting objects by the latter definition. Note that \shark\ galaxies with $S_{\rm band-7}>0.1$~mJy and $S_{\rm band-6}>0.1$~mJy are MS galaxies at $z\lesssim 2$ and only mildly above the MS at higher redshifts, $\rm sSFR/sSFR_{\rm MS}\approx 1.3-2$. In band-4, galaxies with $S_{\rm band-4}>0.1$~mJy are more extreme, with $\rm sSFR/sSFR_{\rm MS}\approx 2-3.5$. Note that one would have to go down to fluxes of $S\approx 0.01$~mJy in ALMA bands 7, 6 and 4 to trace typical MS galaxies across the whole redshift range studied here.
Compared to AS2UDS and ALESS, we find that the predicted sSFR of ALESS-like \shark\ galaxies agrees well with the observations (within the scatter). There is a hint of a trend of sSFR decreasing with increasing redshift in the observations, but is highly uncertain given the large errorbars. \shark\ predicts the opposite: the sSFR of ALMA band selected galaxies increases with increasing redshift at fixed flux, in a way that mimics the evolution of the MS normalisation.
\begin{figure*}
\begin{center}
\includegraphics[trim=18mm 15mm 27mm 27mm, clip,width=0.77\textwidth]{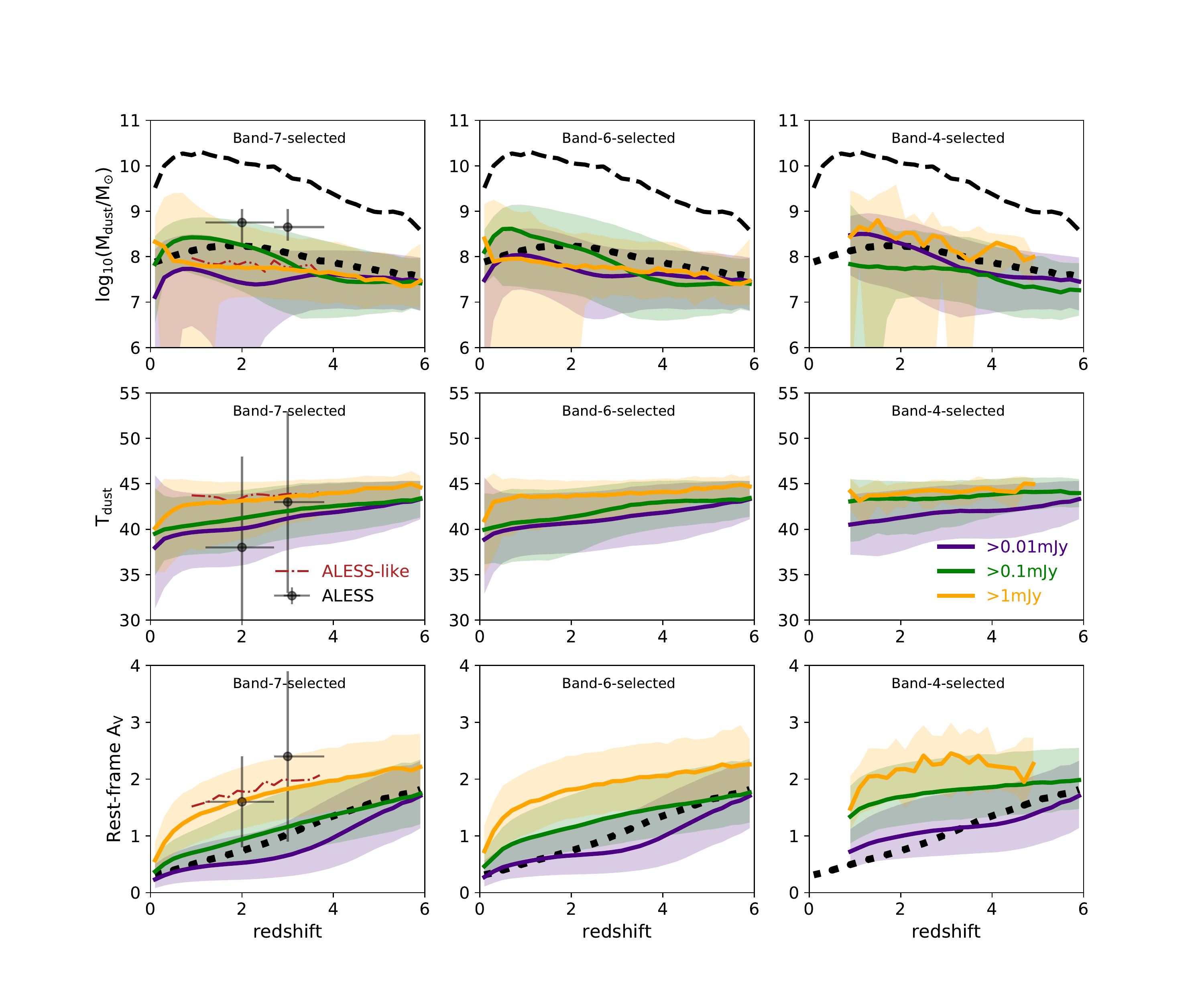}
\caption{As in Fig.~\ref{propssmgs} but for dust  mass, dust temperature and rest-frame attenuation in the V-band. We include the observational constraints of \citet{daCunha15}, which are meant to be comparable to the band-7 sources that are brighter than $1$~mJy. The dotted lines show the median dust mass (top panels) and $A_{\rm V}$ (bottom panels) of the MS galaxies shown in middle panels of Fig.~\ref{propssmgs}; dot-dashed lines in the left hand panels show the median for the \shark\ ALESS-like sample of the bottom panel of Fig.~\ref{smgseds}; and dashed lines in the top panels show the median dust mass of the $100$ most dust massive galaxies.}
\label{propssmgs2}
\end{center}
\end{figure*}
\subsection{The host halo masses of SMGs}

The environment of SMGs has been a topic of great interest in the literature. \citet{Hickox12} measured the spatial clustering of SMGs selected from the $870\mu$m Large APEX Bolometer Camera (LABOCA) instrument, that have fluxes $\gtrsim 4$~mJy, and from the clustering biases inferred the host halo masses to be $\approx 10^{12.8}\,\rm M_{\odot}$. Blending of sources due to the poor angular resolution of LABOCA may decrease this host halo mass by a factor of $\approx 2-6$  to $\approx 10^{12}-10^{12.5}\,\rm M_{\odot}$ \citep{Cowley17b}. 

The bottom panels of Fig.~\ref{propssmgs} show the median host halo mass for galaxies selected in bands 7, 6 and 4 and using three different threshold fluxes. We also show for reference the median halo mass of the $100$ most massive halos as a function of redshift. We generally find that brighter galaxies in the three ALMA bands analysed here inhabit more massive halos. The difference in the median mass for the faintest and brightest flux thresholds increases from band-7 to band-4: galaxies with $S>0.01$~mJy reside in halo masses $\approx 0.8$~dex, $\approx 1.2$ and $\approx 1.5$~dex less massive than galaxies with $S>1$~mJy in bands 7, 6 and 4, respectively. We also find that band-4 selected galaxies reside in more massive halos at fixed flux than band 6 and 7 galaxies. In fact, at $z>3.5$, $S_{\rm band-4}>1$~mJy galaxies are {\it excellent tracers of the most massive halos}, and hence represent a unique probe of proto-clusters, {given the expected correlation between the mass of progenitor halos and their $z=0$ descendants \citep{Muldrew15}. \citet{Casey19} in a blind ALMA band~4 survey detected a bright source,  $S_{\rm band-4}\approx 0.7$~mJy, at $z\approx 5.8$ that is consistent with being a proto-cluster of mass $\approx 10^{12.5}\,\rm M_{\odot}$, providing a tentative confirmation of our prediction.}

The median host halo mass of galaxies with $S>1$~mJy vary little with redshift $\lesssim 0.5$~dex, while the overall evolution of the halo mass function is evolving much more strongly. For reference, at a fixed number density of $10^{-3} \rm h^3\, cMpc^{-3}$, the halo mass evolves by $\approx 1.5$~dex from $\approx 10^{12.4}\rm M_{\odot}\, h^{-1}$ at $z=0$ to $\approx 10^{10.9}\rm M_{\odot}\, h^{-1}$ at $z=6$ (in a $\Lambda$CDM universe). This means that the clustering bias of these ALMA-selected SMGs is increasing with increasing redshift. This is also clear from the way the median halo mass of band-7 and 6 selected galaxies with $S>1$~mJy approaches the mass of the $100$ most massive halos in the lightcone at $z\rightarrow 6$. We show the observational estimate of \citet{Hickox12} in the bottom, left panel of Fig.~\ref{propssmgs} and found it to be consistent with the predicted median halo mass of ALESS-like \shark\ galaxies, within the errorbars. The small difference in the latter comparison can simply be due to the sample of \citet{Hickox12} being brighter than ALESS and/or source confusion \citep{Cowley17b}. \citet{Fujimoto16} computed an upper limit for the typical halo mass hosting $S_{\rm band-6}\gtrsim 0.1$~mJy sources using the counts-in-cell method of $10^{12.5}\rm \,M_{\odot}$ consistent with our predicted median halo mass for those galaxies, $10^{12}\rm \, M_{\odot}$.

\subsection{Dust masses, temperatures and rest-frame optical attenuation of SMGs}

The top and middle panels of Fig.~\ref{propssmgs2} present the dust mass and temperature evolution with redshift for galaxies selected in ALMA bands 7, 6, 4 to have fluxes $>0.01$~mJy, $>0.1$~mJy and $>1$~mJy. 
In bands 6 and 7, the dust mass correlates non-linearly with FIR flux, with the $S>0.1$~mJy sample having the highest dust masses on average, of the three flux threshold samples. The $S>1$~mJy in these bands have on average $\lesssim 0.3$~dex less dust mass than the $S>0.1$~mJy galaxies, and similar to the faint sample $S>0.01$~mJy. This is a direct consequence of the fact that in \shark\ MS galaxies have the highest dust masses at fixed stellar mass (see bottom panels of Fig.~\ref{tempMS}), with the dust mass decreasing as one moves above the MS. However, in band~4 the $S_{\rm band-4}>1$~mJy galaxies also have higher dust masses. This is due to the fact that those galaxies correspond to the most massive, starbursting galaxies in \shark, which also have large amounts of dust (see galaxies above the MS and with $M_{\star}>10^{11}\,\rm M_{\odot}$ in the bottom panels of Fig.~\ref{tempMS}). In any case, neither of these ALMA bands selected samples trace the most dust massive galaxies (see dashed lines).

In the middle panels of Fig.~\ref{propssmgs2} we show the evolution of $T_{\rm dust}$. There is a weak evolution of an increasing $T_{\rm dust}$ with increasing redshift for the faint and intermediate flux samples ($S>0.01$~mJy and $>0.1$~mJy, respectively), while the brightest samples ($S>1$~mJy) display almost no evolution, with $T_{\rm dust}$ being $\approx 42-44$~K at all redshifts. Fig.~\ref{tempMS} shows that $T_{\rm dust}$ varies with both $M_{\star}$ and SFR, but at high $M_{\star}$ and SFRs typical of the galaxies traced in the selections of Fig.~\ref{propssmgs2}, $T_{\rm dust}$ is always $\approx 40-45$~K. This behaviour in \shark\ is due to the underlying assumption that dust in its diffuse and birth cloud phases have two distinct but constant temperatures. Note that the ALESS-like \shark\ sample has dust masses and temperatures consistent with the $S_{\rm band-7}>1$~mJy sample, despite its average brighter flux. The high $T_{\rm dust}$ that we find for SMGs indicate that dimming due to the cosmic microwave background \citep{dacunha13,Jin19} is not expected to have an important effect in our predictions and hence can be neglected.

Compared to the derived physical properties of ALESS galaxies presented in \citet{daCunha15}, we find that \shark\ predicts dust masses that are too low by $0.5-1$~dex, which is compensated by the slightly higher dust temperatures.
In a grey body, the bolometric IR luminosity, $L_{\rm IR}\propto M_{\rm dust}\, T^{4+\beta}_{\rm dust}$, with typical values of $\beta\approx 1-3$. Because the dependence with temperature is much stronger, at fixed luminosity, large variations in dust mass can be compensated by small variations in temperature. Interestingly, the bottom panel of Fig.~\ref{smgseds} shows that the median \shark\ SED of ALESS-like galaxies is very similar to the observed SED reported in \citet{daCunha15}. This suggests that the differences seen in the top and middle panels of Fig.~\ref{propssmgs2} are at least in part due to the systematic effects introduced in fitting the SEDs of galaxies to infer dust masses and temperatures. Inferences of dust temperature in dusty galaxies have not reached consensus, with some other studies indicating $T_{\rm dust}\approx 50$~K to be preferred \citep{Jin19,Cortzen20,Yang20}. {Adopting the latter temperature (instead of the $\approx 40$~K found in \citealt{daCunha15}) and $\beta=1$ and $=3$, one would expect a reduction in the dust mass of a factor $3$ and $4.7$, respectively, in better agreement with our predictions.}

The bottom panels of Fig.~\ref{propssmgs2} show the evolution of the rest-frame $A_{\rm V}$ for \shark\ galaxies selected in ALMA bands 7, 6 and 4. The FIR flux of galaxies is correlated with $A_{\rm V}$ so that a higher ALMA band flux is associated to a higher $A_{\rm V}$ at fixed redshift. In any one of the flux selections, $A_{\rm V}$ increases with increasing redshift. We find that the samples with $S>1$~mJy have $A_{\rm V}>1$ at $z>0.3$ in all the analysed ALMA bands, and on average $A_{\rm V}$ is higher for galaxies selected at longer wavelengths and at fixed flux (i.e. $S_{\rm band-4}>1$~mJy galaxies have a higher $A_{\rm V}$ than $S_{\rm band-7}>1$~mJy at fixed redshift). For reference, the dotted line shows the $A_{\rm V}$ evolution of MS galaxies with $10^9\,\rm M_{\odot}<M_{\star}<10^{10}\,\rm M_{\odot}$. \shark\ galaxies with $S_{\rm band-7}>0.1$~mJy follow closely the $A_{\rm V}$ evolution of MS galaxies of masses $10^9\,\rm M_{\odot}<M_{\star}<10^{10}\,\rm M_{\odot}$. In comparison, band-6 and 4 galaxies selected using the same flux threshold are above the MS $A_{\rm V}$ values.
Compared to the observations, we find that the ALESS-like \shark\ galaxies produce consistent $A_{\rm V}$ values within the uncertainties, and show a qualitatively similar evolutionary trend. This is one of the main reason why the FUV-to-NIR part of the median SED of ALESS-like SMGs agree well with observations (bottom panel of Fig.~\ref{smgseds}).

Together with $A_{\rm V}$ evolving with redshift, the slope of the attenuation curve $\eta$ in Eqs.~\ref{taus_screen}~and~\ref{tau_bc} is also changing. We measure an effective $\eta_{\rm eff}$ following Eq.~\ref{temp_calc}, but instead of temperature we use $\eta_{\rm ISM}$ and $\eta_{\rm BC}$. We find that $\eta_{\rm eff}$ increases from $-0.77$ at $z=0$ to $-0.45$ at $z=6$ for the galaxy sample with $S_{\rm band-7}>1$~mJy. This means that attenuation curves become greyer with increasing redshift, in agreement with the RT analysis of hydrodynamical simulations by \citet{Narayanan18}. 

\begin{figure*}
\begin{center}
\includegraphics[trim=18mm 15mm 27mm 27mm, clip,width=0.77\textwidth]{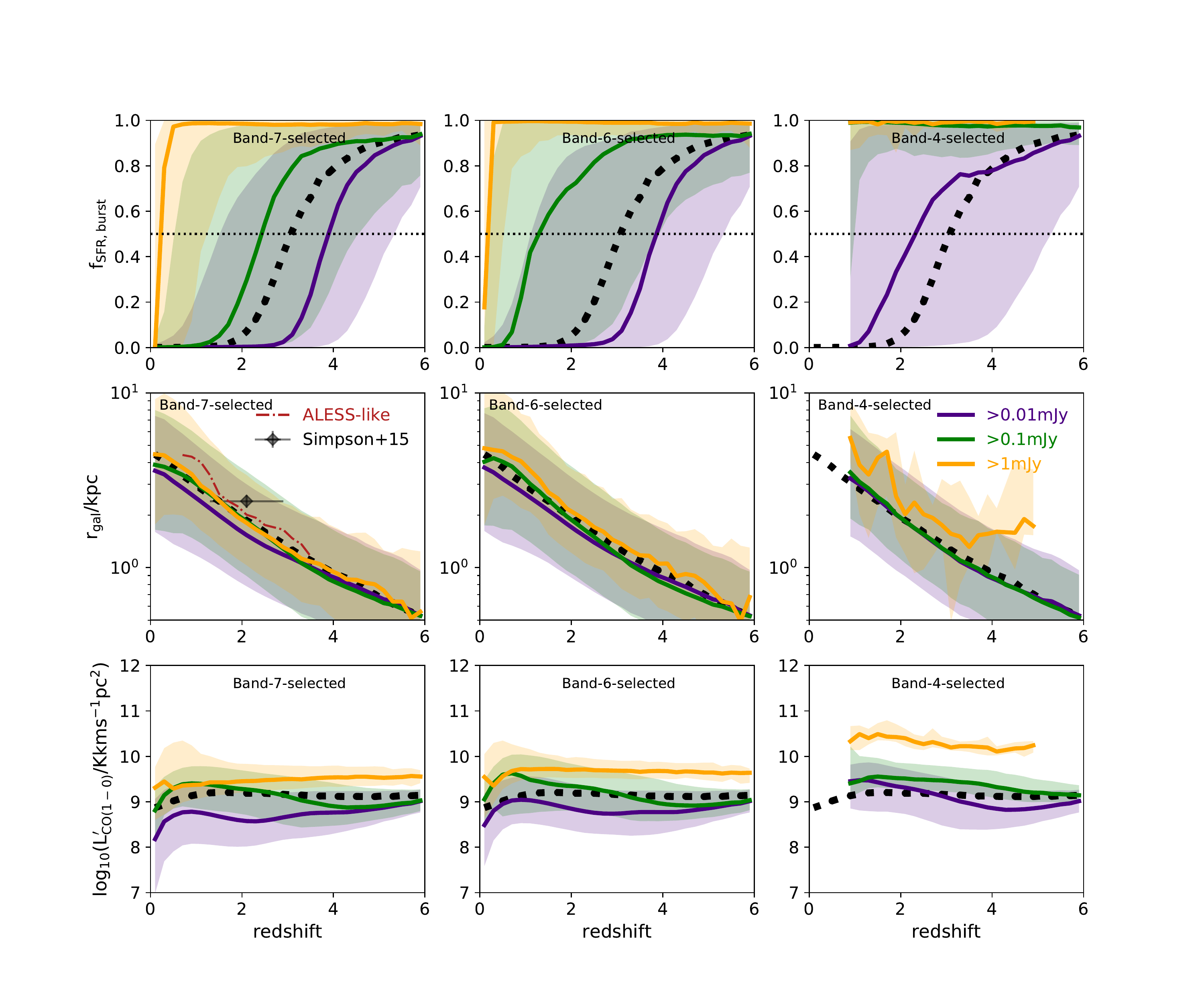}
\caption{As in Fig.~\ref{propssmgs} but for the fraction of SFR that is in the form of a starburst, the physical SFR-weighted effective radius of galaxies, and the CO(1-0) brightness luminosity. In each panel, the dotted lines show the evolution of the median of MS galaxies defined as in Fig.~\ref{propssmgs}. In the middle, left hand panel, we also show the observational estimates of \citet{Simpson15} for the ALESS sample, together with the median of the \shark\ ALESS-like sample of the bottom panel of Fig.~\ref{smgseds}.}
\label{propssmgs3}
\end{center}
\end{figure*}

\subsection{Star formation modes in SMGs}

The top panels of Fig.~\ref{propssmgs3} show the fractional contribution of the burst mode of SF to the total instantaneous SFR of galaxies as a function of redshift for three flux samples selected in bands 7, 6 and 4. We remind the reader that the burst mode of SF can be triggered by both galaxy mergers and global disk instabilities. In both cases gas is driven towards the centre of the galaxy and a starburst takes place from that gas reservoir. The SF law applied in the burst mode is the same as the normal mode of SF, except for a $10$ times shorter molecular gas depletion time (see $\S$~\ref{modeldescription} for a description).

Galaxies with $S>1$~mJy in bands 7, 6 and 4 have SFRs dominated by the burst mode at $z\gtrsim 0.7$, $z\gtrsim 0.4$ and throughout the whole redshift range, respectively. However, from Fig.~\ref{numbercounts} and the bottom panel of Fig.~\ref{smgseds}, we know that this burst mode is split between galaxy merger- and disk instabilities-driven in a ratio close to 50-50.
In the case of galaxies with $S>0.1$~mJy, we find that their SFR is dominated by the burst mode at $z\gtrsim 2.5$, $z\gtrsim 1.5$ and across the whole redshift range for galaxies selected in band 7, 6 and 4, respectively. In the faintest sample $S>0.01$~mJy, we find their SFR to be dominated by the burst mode at $z\gtrsim 4$ for bands 7 and 6 and $z\gtrsim 2.5$ for band 4. For reference, MS galaxies with stellar masses $10^9-10^{10}\,\rm M_{\odot}$ have their SFR dominated by bursts at $z\gtrsim 3$. This means that only pushing down to fluxes of $0.01-0.1$~mJy one is able to get a sample that is representative of {MS galaxies whose sub-mm emission is induced by both starbursts and `normal' SF (see also the middle panel of Fig.~\ref{propssmgs})}. This is extremely challenging even with ALMA. 

\subsection{Galaxy sizes of SMGs}

Observations with ALMA of SMGs have revealed rather compact galaxies (with FIR continuum sizes $1-5$~kpc; e.g. see  \citealt{Ikarashi15,Hodge16,Oteo17,Fujimoto17}). However, SMGs do not appear necessarily more compact than high-z MS galaxies \citep{Elbaz18,Gomez-Guijarro19,Puglisi19}, and it appears like these sizes do not depend on merger state \citep{Fujimoto17}. \citet{Fujimoto17} found a weak trend of galaxy sizes to increase with increasing FIR luminosity, with galaxies of $L_{\rm FIR}\approx 10^{12}\,\rm L_{\odot}$ and $L_{\rm FIR}\approx 10^{13}\,\rm L_{\odot}$ having effective radii $\approx 1$~kpc and $\approx 2$~kpc, respectively. Here, we explore the SFR-weighted effective radii, $r_{\rm gal}$, of galaxies in \shark\ selected using three different flux thresholds in ALMA bands 7, 6 and 4. These are presented in the middle panels of Fig.~\ref{propssmgs3}. {We compute $r_{\rm gal} = (\rm SFR_{\rm disk}\times r_{\rm 50,disk} + SFR_{\rm bulge} \times r_{\rm 50,bulge})/ SFR_{\rm total}$, where $\rm SFR_{\rm disk}$ and $\rm SFR_{\rm bulge}$ are the SFRs contributed by the disk and bulge, respectively, to $\rm SFR_{\rm total}$, and $r_{\rm 50,disk}$ and $r_{\rm 50,bulge}$ are the half-gas mass radii of the disk and bulge, respectively.}

Overall, we find that $r_{\rm gal}$ decreases with increasing redshift at fixed flux and a weak dependence of $r_{\rm gal}$ on the observed flux at fixed redshift, so that brighter galaxies in the submm have on average larger $r_{\rm gal}$, in qualitative agreement with \citet{Fujimoto17}. However, the latter is very weak and almost completely disappears for galaxies at $z\gtrsim 3$. The only galaxy sample that is clearly more extended throughout the whole redshift range is that with $S_{\rm band-4}>1$~mJy. For reference, we also show the size evolution of MS galaxies with stellar masses $10^9-10^{10}\,\rm M_{\odot}$ and find that those are indistinguishable from band 7 and 6 bright galaxies, $S>1$~mJy, in agreement with the conclusions in \citet{Barro16,Elbaz18}.

The sizes of galaxies with $S>1$~mJy in bands 7 and 6 (which are the bands used for the observational studies above) are within the observed ones, with typical values of $1-3$~kpc. We also show the \citet{Simpson15} {FWHM} observational estimate for the ALESS sample, which should be compared to the median of the ALESS-like \shark\ sample (dot-dashed line in the middle, left panel of Fig.~\ref{propssmgs3}). The ALESS-like \shark\ sample is slightly more extended than the $S_{\rm band-7}>1$~mJy, reflecting the weak correlation we find between the median galaxy size and the band-7 flux. Within the uncertainties, we find \shark\ to agree well with \citet{Simpson15}.

\subsection{The molecular gas emission of SMGs}\label{cosec}

\begin{figure}
\begin{center}
\includegraphics[trim=0mm 2.8mm 7.5mm 10mm, clip,width=0.48\textwidth]{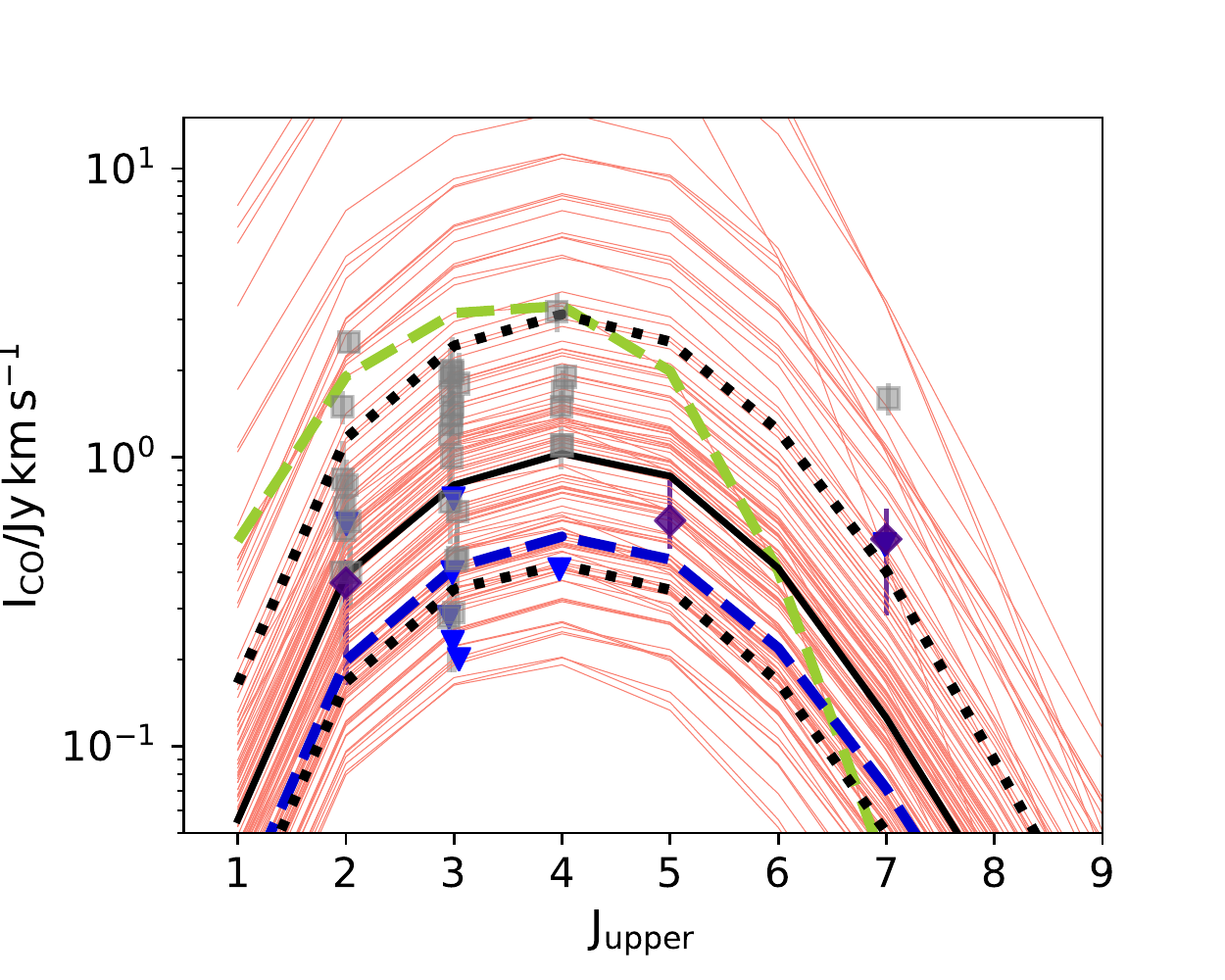}
\caption{CO SLED for SMGs selected in \shark\ to follow the $850\mu$m flux distribution of the SMG sample of \citet{Bothwell13}. Thin lines show individual \shark-selected galaxies, while the thick solid and dotted lines show the median and the $1\sigma$ percentile ranges. The green and blue dashed lines show examples of SMGs with very weak and strong AGN hard X-ray emission, respectively.
Individual detections from \citet{Bothwell13} are shown as squares, while upper limits are shown as down-pointing triangles. The x-axis position of observations are slightly perturbed to aid visualization. We also show the median CO SLED for galaxies with $L_{\rm IR}>10^{12}\,\rm L_{\odot}$ above the MS by a factor $\ge 3.5$, equivalent to the galaxies in \citet{Bothwell13}, presented in \citet{Valentino20b} (diamonds), which we place at $z=2$ for comparison with \shark\ and \citet{Bothwell13} (their original median redshift is $z=1.3$).}
\label{cosleds}
\end{center}
\end{figure}
\begin{figure*}
\begin{center}
\includegraphics[trim=18mm 10mm 27mm 5mm, clip,width=0.95\textwidth]{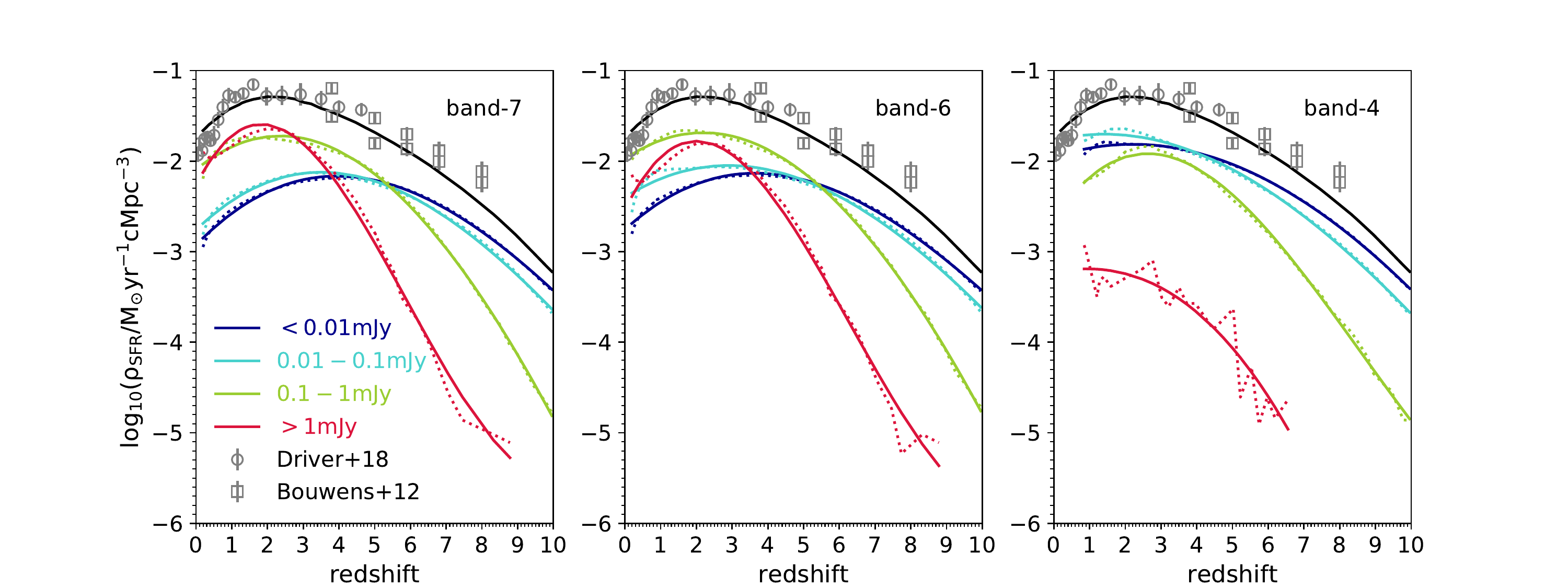}
\includegraphics[trim=17mm 0mm 28mm 5mm, clip,width=0.95\textwidth]{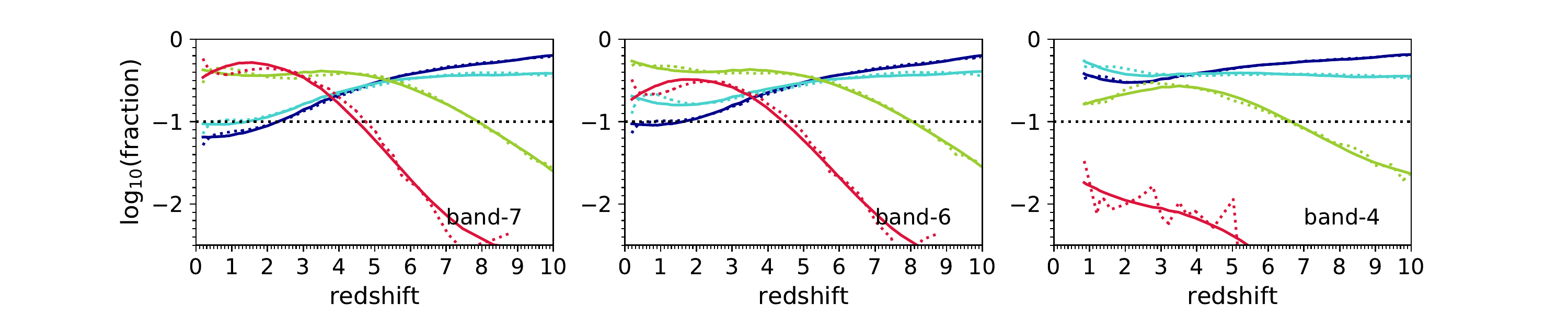}
\caption{{\it Top panel:} Cosmic SFR of the universe. The black line shows the predicted total cosmic SFR, while data points show the observational estimates of \citet{Driver17} and \citet{Bouwens12}. For the latter we show both the corrected and uncorrected measurements, which were derived from UV LFs.
The colour dotted lines show the predicted SFR density contribution of galaxies selected in ALMA band-7 (left panel), band-6 (middle panel) and band-4 (right panel) selected galaxies in $4$ different flux bins. The coloured solid lines show a spline fit to the latter, which are presented in Table~\ref{TableSplineFits}. {\it Bottom panel:} The fractional contribution from the samples of the top panel to the total SFR density predicted by \shark. For reference, the horizontal dotted line shows a contribution of $10$\%.}
\label{SFRevo}
\end{center}
\end{figure*}

The bottom panels of Fig.~\ref{propssmgs3} shows the CO(1-0) brightness luminosity as a function of redshift, $L^{\prime}_{\rm CO(1-0)}$, for three samples selected in bands 7, 6 and 4. We find that higher fluxes are associated with higher $L^{\prime}_{\rm CO(1-0)}$, and that at fixed flux, galaxies selected in band 4 have higher $L^{\prime}_{\rm CO(1-0)}$ than those selected in band 6 and 7. For reference we also show the predicted $L^{\prime}_{\rm CO(1-0)}$ of MS galaxies with stellar masses $10^9-10^{10}\,\rm M_{\odot}$ and find the samples with $S>1$~mJy in bands 7, 6 and 4 and the $S_{\rm band-4}>0.1$~mJy sample to be CO(1-0) brighter. However, note that the samples with $S>1$~mJy in bands 7 and 4 are only $\approx 0.3-0.5$~dex above the median $L^{\prime}_{\rm CO(1-0)}$ of MS galaxies, while galaxies with $S_{\rm band-4}>1$~mJy are $\approx 1.2$~dex above MS galaxies. Some of these differences come from the different stellar masses being probed by the samples (see top panels in Fig.~\ref{propssmgs}), given that $L^{\prime}_{\rm CO(1-0)}$ increases with increasing stellar mass, on average, and some of this is due to the complex scaling between $L^{\prime}_{\rm CO(1-0)}$ and SFR at fixed stellar mass presented in Fig.~\ref{tempMSCO}.

\citet{Bothwell13} presented one of the most comprehensive studies of the CO SLED of SMGs in the literature, which has been extended to brighter \citep{Spilker14,Yang17,Canameras18} and fainter \citep{Valentino20b} objects. In order to compare with \citet{Bothwell13}, we select galaxies in \shark\ to follow the same $850\mu$m continuum emission of the sample in \citet{Bothwell13} as it was done with the ALESS-like sample of Fig.~\ref{smgseds}. The median redshift of this \shark\ Bothwell13-like sample is $2.15$, while for \citet{Bothwell13} this is $2.28$, in excellent agreement.

From the Bothwell13-like sample, we compute the median CO SLED with the corresponding $16^{\rm th}$-$84^{\rm th}$ percentiles ranges and compare with the observations in Fig.~\ref{cosleds}. 
Fig.~\ref{cosleds} also shows the individual \shark\ galaxies in the Bothwell13-like sample. \shark\ appears to be systematically fainter than the \citet{Bothwell13} observations, with detectiosn falling preferentially between the median and the $84^{\rm th}$ percentile of \shark\ galaxies.

Note that although most CO SLEDs in Fig.~\ref{cosleds} peak at around $J_{\rm upper}=4-5$, there is great diversity in their shapes. This is most noticeable at $J_{\rm upper}\ge 5$, where some galaxies experience a sharp decrease in the CO intensity, while others display a gentle decrease. The latter is driven by the effect of AGN in those SMGs, modelled by the inclusion of the hard X-ray flux (see $\S$~\ref{COmodel}), which tends to boost the line intensity of high CO transitions. An example of this is visualised in Fig.~\ref{cosleds}, where we highlight two SMGs with a low ($\approx 10^{38}\,\rm erg\, s^{-1}$) and high ($\approx 2\times 10^{43} \,\rm erg\, s^{-1}$) AGN hard X-ray luminosity. The former displays a sharp $I_{\rm CO}$ decrease at $J_{\rm upper}>4$, while the latter displays a gentle decay. The reason why the X-ray bright galaxy has a lower $I_{\rm CO}$ at $J_{\rm upper}<5$ is because it is at higher redshift, $z\approx 2.3$, compared to the faint one, $z\approx 1.7$.

To compare \shark\ with a more representative median CO SLED of bright starburst galaxies, we show in Fig.~\ref{cosleds} the measurements of \citet{Valentino20b} for starburst galaxies with $L_{\rm IR}>10^{12}\rm \, L_{\odot}$ that are above the MS by factors $\ge 3.5$, similar to the galaxies in \citet{Bothwell13}. We find the \shark\ $I_{\rm CO}$ medians to be in excellent agreement with \citet{Valentino20b} at transitions $J_{\rm upper}=2,\, 5$, but are too faint at $J_{\rm upper}=7$. This shows that despite the CO SLED modelling included in \shark\ accounting for an X-ray like source, its effect is likely insufficient to explain the high excitation at high $J_{\rm upper}$. Inclusion of additional heating sources, such as shocks, could increase the excitation at higher $J_{\rm upper}$. 

It is worth noting that the level of agreement we find with observed CO SLEDs is somewhat surprising given the assumptions we make in our modelling (see $\S$~\ref{COmodel} for a discussion), and shows that some of the effects we ignore may be cancelling each other to some extent.

\section{The contribution of SMGs to the Cosmic Star Formation Rate and H$_2$ density History}\label{sec:CSFRD}

\begin{figure*}
\begin{center}
\includegraphics[trim=18mm 10mm 27mm 5mm, clip,width=0.95\textwidth]{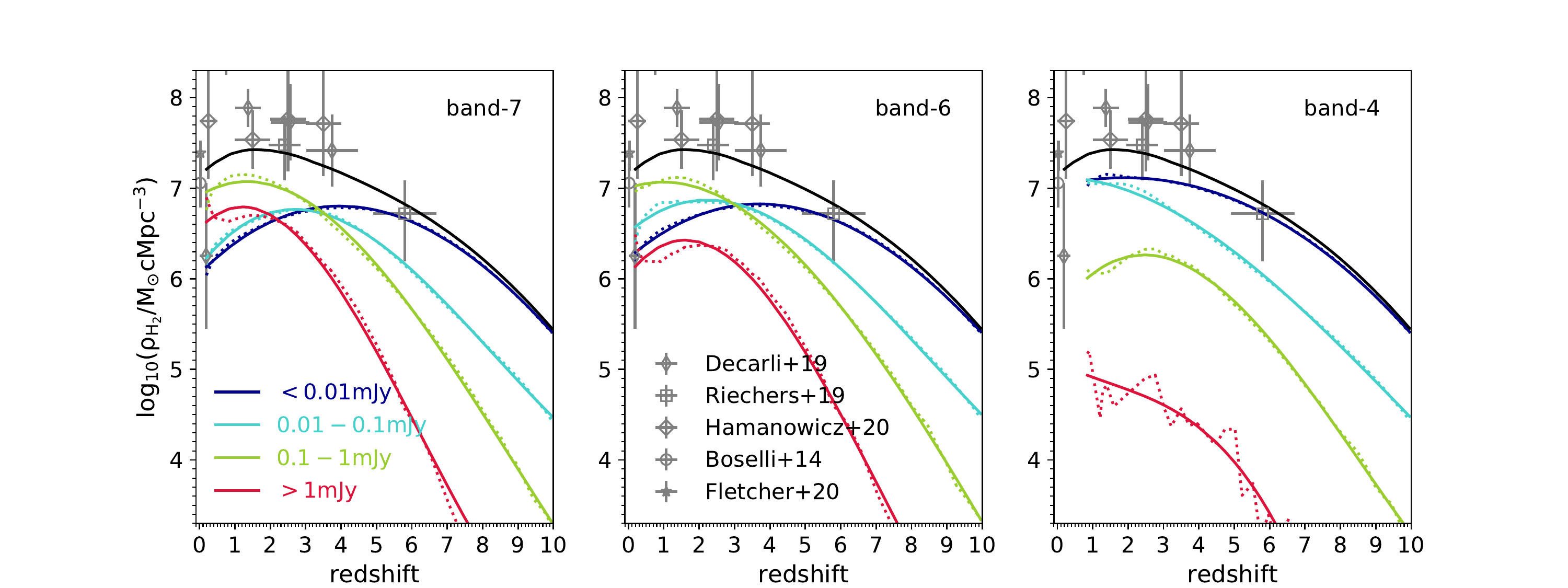}
\includegraphics[trim=17mm 0mm 28mm 5mm, clip,width=0.95\textwidth]{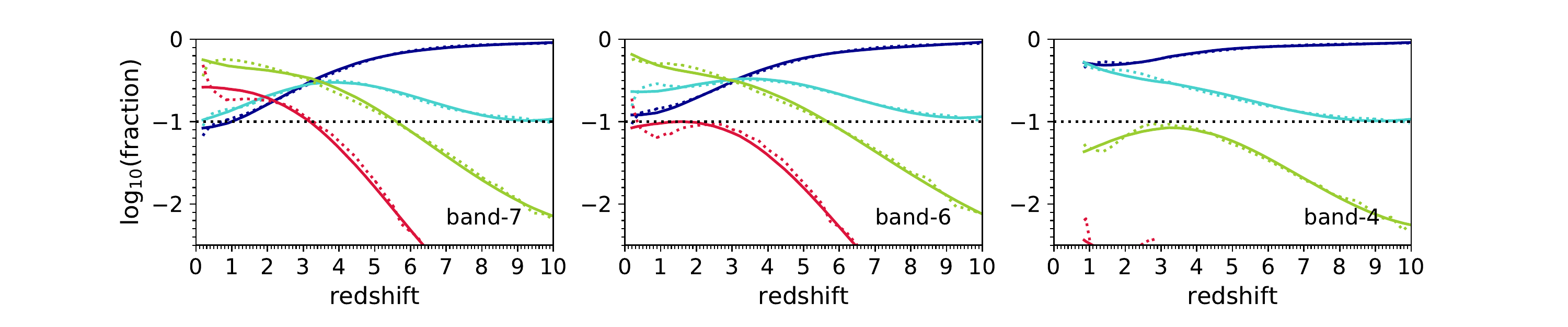}
\caption{As in Fig.~\ref{SFRevo} but for the cosmic H$_2$ density evolution of the universe. Observational estimates shown here correspond to \citet{Boselli14,Fletcher20,Decarli19,Riechers19}; Hamanowicz et al. (in preparation), as labelled in the middle panel.}
\label{H2evo}
\end{center}
\end{figure*}

Current observational estimates of the contribution of FIR-selected galaxies to the CSFRD of the Universe are sparse at $z\gtrsim 4$, due to the small areas probed by deep FIR surveys \citep{Casey18b}. The success of \shark\ in reproducing a variety of observed properties of SMGs, from their optical-to-NIR emission, to their derived intrinsic properties, makes it an ideal tool to explore this question. 

Fig.~\ref{SFRevo} shows the CSFRD evolution at $0\le z\le 10$ and the contribution from galaxies selected in different flux ranges in ALMA bands 7, 6 and 4. The total CSFRD is the same in all panels. We show observational constraints from \citet{Driver17,Bouwens12}. For the latter we show the uncorrected and corrected measurements, which were derived from rest-frame UV LFs. From the predicted total CSFRD in \shark\ we can conclude that rest-frame UV-selected galaxies after being corrected by dust attenuation, are able to recover the total CSFRD closely. Hence, in our model framework, no significant cosmic SFR would be missing, which would agree with an overall dust-poor galaxy population at $z\gtrsim 5$ (similar to ``model A'' in \citealt{Casey18a}). This is because most galaxies in \shark\ at those high redshifts are metal poor ($Z_{\rm gas}<0.25\,Z_{\odot}$), and only a handful of galaxies at those redshifts have metallicities above that (see Fig.~$14$ in \citealt{Lagos19} and analysis therein). 

Moving to the contribution from different ALMA band selected galaxies to the total CSFRD, we can see that galaxies with $S_{\rm band-7}>1$~mJy contribute a constant $\approx 44$\% at $0\le z\le 3$, after which their contribution starts to decrease. Nevertheless, $S_{\rm band-7}>1$~mJy galaxies should make up $>10$\% of the total CSFRD out to $z \approx 5$, well within the ``cosmic dawn'' regime. Going fainter to $0.1<S_{\rm band-7}/\rm mJy<1$, we see that these galaxies contribute $>10$\% at $z\lesssim 8$. Fainter galaxies, however, dominate at $z\gtrsim 6$.
\shark\ galaxies selected in ALMA band 6 behave similarly, though the contribution of galaxies with $S_{\rm band-6}>1$~mJy never exceed $\approx 28$\%.
Band 4 selected galaxies behave quite differently, with  $S_{\rm band-4}>1$~mJy making a negligible contribution to the CSFRD at all redshifts. However, going to galaxies with $0.1<S_{\rm band-4}/\rm mJy<1$, we find that they make up $>10$\% of the CSFRD at $z<6.8$. This means that $2$~mm selected surveys should aim to go down to fluxes of $0.1$~mJy if the aim is to constrain the contribution of dust galaxies to the CSFRD. 

Another quantity that has sparked significant attention with the advent of ALMA is the cosmic molecular gas density, $\rho_{\rm H_2}$, evolution. Recent constraints from the ASPECS survey \citep{Decarli19}, COLDz \citep{Riechers19} and ALMACAL-CO (Hamanowicz et al. in preparation) suggest $\rho_{\rm H_2}$ increases from $z=0$ to $z\approx 1.5-3.5$ followed by a decline towards higher redshifts. This is qualitatively similar to the CSFRD, however, the exact redshift peak of $\rho_{\rm H_2}$ and the magnitude of its evolution are still highly uncertain. We use \shark\ to investigate the contribution from the flux selected samples above to $\rho_{\rm H_2}$ as a function of redshift in Fig.~\ref{H2evo}. First, comparing the total $\rho_{\rm H_2}$ to observations, we find our measurements agree very well with the measurements within the uncertainties, except for the ASPECS $\rho_{\rm H_2}$ at $z\approx 1.5$. There is a clear tension between the ASPECS and the COLDz plus ALMACAL-CO measurements at $z\approx 1.5-2$, which likely comes from the underlying assumed CO SLED of the surveys being different. \citet{Riechers20} argued that ASPECS assumes a CO SLED that were too low excitation based on new measurements of CO(1-0). If that is the case, then the \citet{Decarli19} measurements are overestimated by a factor of $\approx 2$. The tension between ASPECS and other $z\approx 0$ measurements on the other hand is most likely due to the small area probed by ASPECS ($0.0013$~deg$^2$), {which implies a cosmic variance of $\approx 50$\% ($78$\%) at $0<z<0.5$ ($0<z<0.25$) according to the cosmic variance calculator of \citet{Driver10}}. This shows that the errorbars reported in observations are likely underestimated as they do not include systematic effects such as the ones mentioned here. 

Regarding the contribution of the different ALMA band flux selected samples, we generally find that bright ALMA bands 7, 6 and 4 galaxies, $S>1$~mJy, contribute less to $\rho_{\rm H_2}$ than their contribution to the CSFRD. The maximal contribution of $S_{\rm band-7}>1$~mJy galaxies to $\rho_{\rm H_2}$ is $\approx 30$\% at $z\approx 0$, decreasing to $<10$\% at $z\gtrsim 3.2$. In the case of $S_{\rm band-6}>1$~mJy galaxies, they make up $\approx 8-10$\% of $\rho_{\rm H_2}$ at $0\lesssim z\lesssim 3.5$, percentage that decreases steeply at higher redshifts. $S_{\rm band-4}>1$~mJy galaxies have a negligible contribution to $\rho_{\rm H_2}$, $\ll 1$\%, throughout the whole redshift range. 

Moving to the samples with $0.1<S/\rm mJy<1$, we find them to be the dominant source of H$_2$ at $z\lesssim 3$ for band 7 and 6. However, at band 4, those galaxies still make up $<10$\% of   $\rho_{\rm H_2}$ at all redshifts. 
It is interesting to see that the fainter sample, $0.01<S/\rm mJy<0.1$ is not dominant at any redshifts in bands 7 and 6, and at the redshift in which $0.1<S/\rm mJy<1$ galaxies cease to be dominant ($z\approx 3$), the majority of H$_2$ starts to come from the faintest galaxies $S<0.01$~mJy. In band 4, we see that galaxies with $S<0.01$~mJy and $0.01<S/\rm mJy<0.1$ contribute similarly to $\rho_{\rm H_2}$ at $z\lesssim 1$, but at higher redshift the vast majority of H$_2$ is contributed by galaxies with $S<0.01$~mJy.

We do not investigate the equivalent atomic hydrogen density here as previous galaxy formation simulations have shown this to be dominated by lower stellar mass galaxies, with a negligible contribution from massive galaxies, $>10^{10}\,\rm M_{\odot}$ (except at low redshift, $z\lesssim 0.2$; e.g. \citealt{Dave13,Lagos14b}). The latter are the typical masses we find in our bright $S>1$~mJy samples (see top panels in Fig.~\ref{propssmgs}).

\section{Discussion and Conclusions}\label{conclusions}

In this work we present a thorough exploration of the properties of submm and mm-selected galaxies in the \shark\ semi-analytic model of galaxy formation across a wide redshift range, $0\le z\le 10$. To predict the broad-band SEDs of galaxies we use \viperfish\ \citep{Lagos19,Robotham20}, which uses the stellar populations of BC03 (with our assumed universal IMF of \citealt{Chabrier03}) and adopts the \citet{Charlot00} attenuation parametrisation. We scale the latter parameters with the galaxy's dust surface density, following the RT analysis of the {\sc EAGLE} hydrodynamical simulations of \citet{Trayford19}. Critical for this work, we re-emit the absorbed light assuming energy balance and adopting 
the IR templates of \citet{Dale14}. We assume two distinct power-law indices for the local interstellar radiation field of the diffuse and birth clouds dust components. This is equivalent to assuming two average dust temperatures for the diffuse and birth cloud dust components that are constant with redshift and galaxy properties. Despite the simplicity of some of these assumptions, we find excellent agreement with observations over a broad range of diagnostics. 
We summarise our main findings below:

\begin{itemize}
    \item We tested the \shark\ predicted number counts against measurements at $650\mu$m, $870\mu$m, $1.1$mm and $2$mm (all the ALMA bands that have been used to compute number counts) finding excellent agreement (Fig.~\ref{numbercounts}). Significant tension is however seen at the $1$mm faint end between observations, and \shark\ predictions fall in between these measurements, though closer to \citet{Fujimoto16}. We also compare the predicted redshift distributions of \shark\ with observations at $100\mu$m, $250\mu$m, $450\mu$m, $850\mu$m, $1.1$mm and $2.2$mm, finding broad agreement (Fig.~\ref{zdist}). There is a small tendency of \shark\ galaxies to be at slightly lower redshifts than some of the most recent $870\mu$m measurements, but given the associated uncertainties of photometric redshifts it is unclear whether this tension is significant. We find that $S>1$~mJy sources are a mix of galaxy merger- and disk instabilities- driven starbursts, showing that they are an inhomogeneous population.
    \item We find that current deep optical surveys (e.g. HUDF and HSC) are deep enough as to detect most of the ALMA band-7 $>1$~mJy SMGs at $z<4$, but are not sensitive enough to find counterparts for all SMGs at higher redshifts (Fig.~\ref{ncountsband7selec}). In fact, in \shark\ $10$\% of these SMGs are IRAC dark in agreement with \citet{Dudzeviciute20}. The latter in \shark\ are at significantly higher redshift, $z\approx 3.5$, compare to the IRAC bright ones, $z\approx 2.2$ (Table~\ref{Iracgals}), and have higher rest-frame V-band attenuation ($A_{\rm V}\approx 2.2$ compared to $A_{\rm V}\approx 1.5$). We find that a nominal 10,000s NIR JWST survey should be deep enough as to see all band-7 $\gtrsim 0.01$~mJy. If we study galaxies selected at bands 6 and 4 we find their optical counterparts to be brighter than band 7 galaxies at fixed flux, so current optical surveys can detect those up to higher redshifts (Figs.~\ref{ncountsband6selec}~and~\ref{ncountsband4selec}). We find that about $40-50$\% of \shark\ galaxies classified as passive in the (u-r) vs. (r-J) plane at $1.5\le z\le 3.5$ are in fact star-forming with $850\mu$m fluxes $>0.1$~mJy (Fig.~\ref{smgcolours}). We compared the stacked FUV-to-FIR SED of \shark\ SMGs that follow the same band 7 flux distribution as ALESS \citet{daCunha15} and compare with their stacked SED finding excellent agreement (Fig.~\ref{smgseds}). All this evidence shows that \shark\ captures the UV-to-MIR properties of SMGs quite well.
    \item Regarding intrinsic properties of submm and mm-selected \shark\ galaxies, we find that  galaxies with $870\mu$m fluxes $>1$~mJy have stellar masses $\approx 10^{10}-10^{11}\,\rm M_{\odot}$, are mildly above the MS by factors of $\approx 2-3$, and live in intermediate mass halos $\approx 10^{12.3}-10^{13}\,\rm M_{\odot}$, in agreement with observations (Fig.~\ref{propssmgs}). These properties are similar to those found in the {\sc EAGLE} hydrodynamical simulations for similarly bright SMGs at $1\le z\le 3$ \citep{McAlpine19}. Interestingly, we find that band~4 galaxies with fluxes $>1$~mJy at $z\gtrsim 4$ in \shark\ trace the most massive galaxies in the simulation, $M_{\star}\gtrsim 10^{11}\rm \,M_{\odot}$, have sSFRs above the MS by factors of $\gtrsim 7-10$ and live in the most massive haloes at those redshifts. This makes them ideal tracers of proto-clusters. We also find that the vast majority of bright SMGs in bands 7, 6 and 4 in \shark\ are experiencing starbursts, but the mechanism behind comes in two very different flavours: galaxy mergers and disk instabilities, almost at the $50-50$ level (Fig.~\ref{propssmgs3}). We also find SMGs to have SFR-weighted effective radii $\approx 1-5$~kpc, with a significant redshift dependence, in a way that high-z SMGs are more compact than low-z SMGs. We also find a very weak positive correlation between the flux and the spatial extent of galaxies at fixed redshift, in agreement with the observational derivations of \citet{Fujimoto17}.
    \item The dust masses and temperatures of \shark\ galaxies with $870\mu$m fluxes $>1$~mJy span a wide range $10^{7}-10^{9}\,\rm M_{\odot}$ and $37-45$~K, respectively (Fig.~\ref{propssmgs2}). These values agree with the observations within the (large) uncertainties. We find that galaxies with $2$mm fluxes $>1$~mJy tend to be more dust rich by $0.5$~dex and have hotter dust temperatures $T_{\rm dust}\approx 42-45$~K than the above sample. We also find that the rest frame V-band attenuation increases with increasing redshift at fixed flux, and for \shark\ galaxies with $870\mu$m fluxes $>1$~mJy the median goes from $A_{\rm V}\approx 1.2$ at $z\approx 1$ to $A_{\rm V}\approx 2$ at $z\approx 4$. These $A_{\rm V}$ values again agree well with observations within the uncertainties. \shark\ galaxies selected at longer wavelengths tend to display higher $A_{\rm V}$ at fixed flux and redshift. In fact, galaxies with $2$mm fluxes $>1$~mJy have $A_{\rm V}\gtrsim 1.8$ at all redshifts (Fig.~\ref{propssmgs2}). 
    \item We study the CO emission of submm and mm-selected galaxies and find that the CO(1-0) luminosity is relatively constant with redshift at fixed flux. Galaxies with $870\mu$m, $1.1$mm and $2$mm fluxes $>1$~mJy have average $L^{\prime}_{\rm CO(1-0)}\approx 10^{9.5}$, $10^{9.7}$ and $10^{10.5}\rm K\,km\,s^{-1}\,pc^{2}$, respectively (Fig.~\ref{propssmgs3}). We build the CO SLED of $850\mu$m SMGs selected to have the same flux distribution as \citet{Bothwell13} and compare to their CO SLEDs. We find that the CO SLED shape resembles the observations, displaying a peak at $J_{\rm upper}\approx 4$, but the normalisation appears to be $\approx 1.5$ times lower in \shark\ (Fig.~\ref{cosleds}). Compared to the median CO SLEDs of starburst galaxies with $L_{\rm IR}>10^{12}\,\rm L_{\odot}$ of \citet{Valentino20b}, we find \shark\ is in excellent agreement at $J_{\rm upper} \le 5$, but is too faint by $\approx 0.2\,\rm Jy\, km\, s^{-1}$ at $J_{\rm upper}=7$. The CO SLED modelling of \shark\ includes an X-ray like source (powered by AGN) in addition to the traditional PDR. Our comparison with observations suggests that additional excitation sources, such as shocks, may be necessary.
    \item Finally, we quantify the contribution of bright $870\mu$m and $1.1$mm-selected galaxies ($>1$~mJy) to the CSFRD (Fig.~\ref{SFRevo}), finding that they make a significant contribution ($>10$\%) at $z\approx 4-5$, but become negligible at higher redshifts. In contrast, $2$mm bright galaxies ($>1$~mJy) make a small contribution, $\lesssim 2$\%, at all redshifts. Our predictions are that dusty galaxies make a negligible contribution to the CSFRD at $z\gtrsim 4$. We extend this analysis to $\rho_{\rm H_2}$ (Fig.~\ref{H2evo}) and find that the contribution of these bright submm and mm galaxies ($>1$~mJy) is even smaller, with $870\mu$m and $1.1$mm-selected galaxies contributing $\gtrsim 10$\% at $z\lesssim 3$ and $\lesssim 2.5$, respectively.
\end{itemize}

The broad agreement we find between the predicted properties of submm and mm-selected galaxies in \shark\ with several observations are very encouraging, and make \shark\ an ideal tool to plan survey strategies to follow up SMGs with different current and upcoming telescopes, and to make predictions about what future surveys may see (which we do here). It also shows that the adopted assumptions in the model are at least appropriate to aid this success, despite their simplicity. 
In the future we will explore physical ways of scaling the dust temperature and radiation field spectrum with local galaxy properties, aided by recent progress in detailed RT analysis of galaxies in hydrodynamical simulations (e.g. \citealt{Trayford19,Lovell20}); and will test the systematic uncertainties of FIR SED fitting to derive dust parameters, such as temperature and mass.

\section*{Acknowledgements}

We thank Chris Lovell, Paul van de Werf, Jacqueline Hogde, Matus Reykn, Joop Schaye, Desika Narayanan and Ian Smail for useful discussions throughout the writing of this manuscript. We also thank the anonymous referee for their insightful report.
CL and EdC have received funding from the ARC Centre of
Excellence for All Sky Astrophysics in 3 Dimensions (ASTRO 3D), through project number CE170100013.
CL also thanks the MERAC Foundation for a Postdoctoral Research Award. 
FV acknowledges support from the Carlsberg Foundation Research Grant CF18-0388 ``Galaxies: Rise and Death''.
GEM acknowledges the Villum Fonden research grant 13160 ``Gas to stars, stars to dust: tracing star formation across cosmic time'' and the Cosmic Dawn Center of Excellence funded by the Danish National Research Foundation under the grant No. 140. 
This work was supported by resources provided by The Pawsey Supercomputing Centre with funding from the
Australian Government and the Government of Western Australia.

\section*{Data Availability}

The lightcone used for this paper is available upon request (please email first author to get access). The SURFS simulation used in this work can be accessed freely from {\url{https://tinyurl.com/y6ql46d4}}. \shark, \viperfish, and {\sc Stingray} are all public codes (see $\S$~\ref{sec:model} and ~\ref{overallstats}), and the python scripts used to produce the plots in this paper can be found at \url{https://github.com/cdplagos/lightcone-analysis}.





\bibliographystyle{mn2e_trunc8}
\bibliography{SHArkIntro}



\appendix

\section{Optical, NIR and MIR counterparts for band-6 and band-4 selected galaxies}

\begin{figure*}
\begin{center}
\includegraphics[trim=20mm 15mm 27mm 22mm, clip,width=0.75\textwidth]{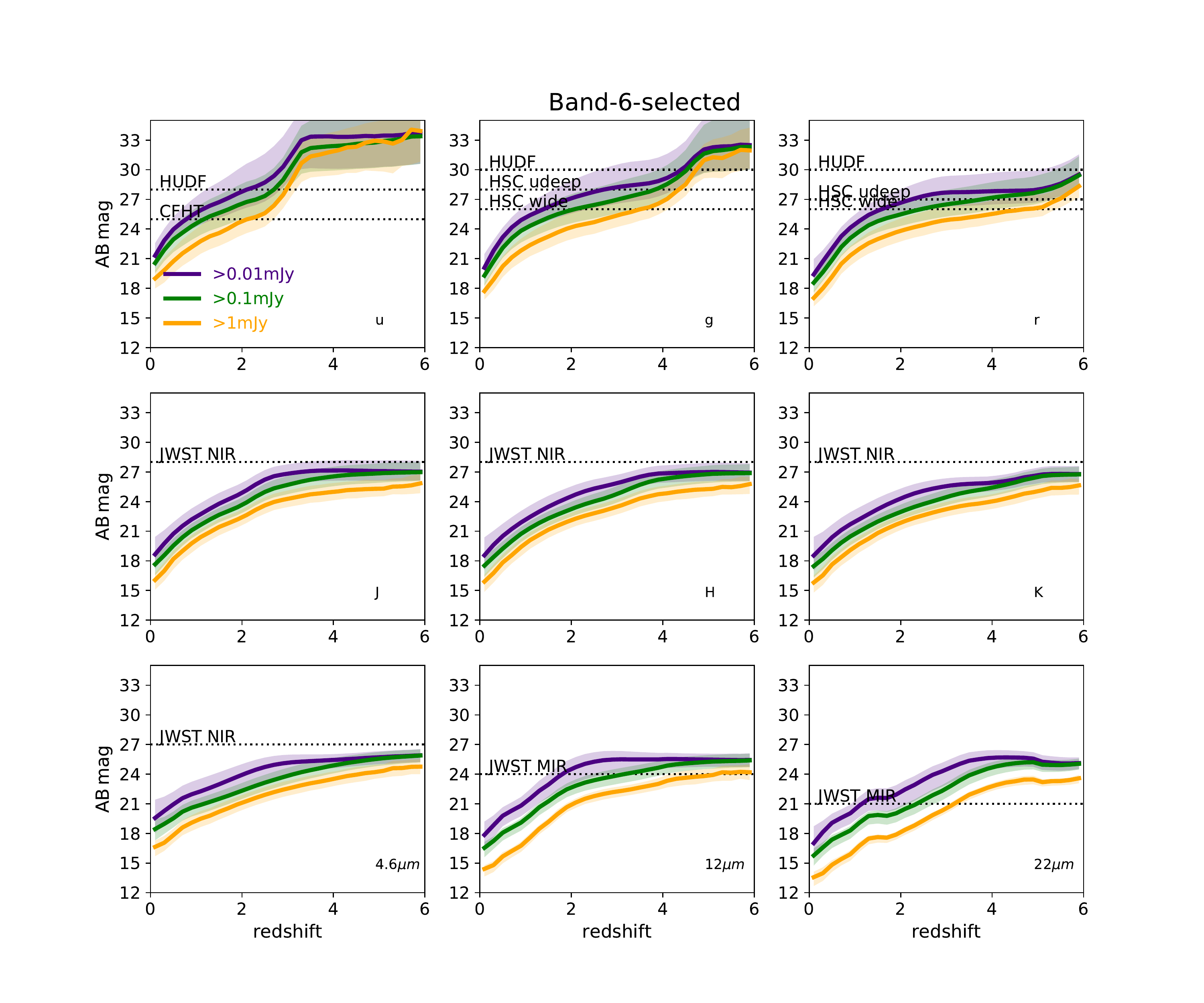}
\caption{As Fig.~\ref{ncountsband7selec} but for ALMA band-6 selected galaxies.}
\label{ncountsband6selec}
\end{center}
\end{figure*}

\begin{figure*}
\begin{center}
\includegraphics[trim=20mm 15mm 27mm 22mm, clip,width=0.75\textwidth]{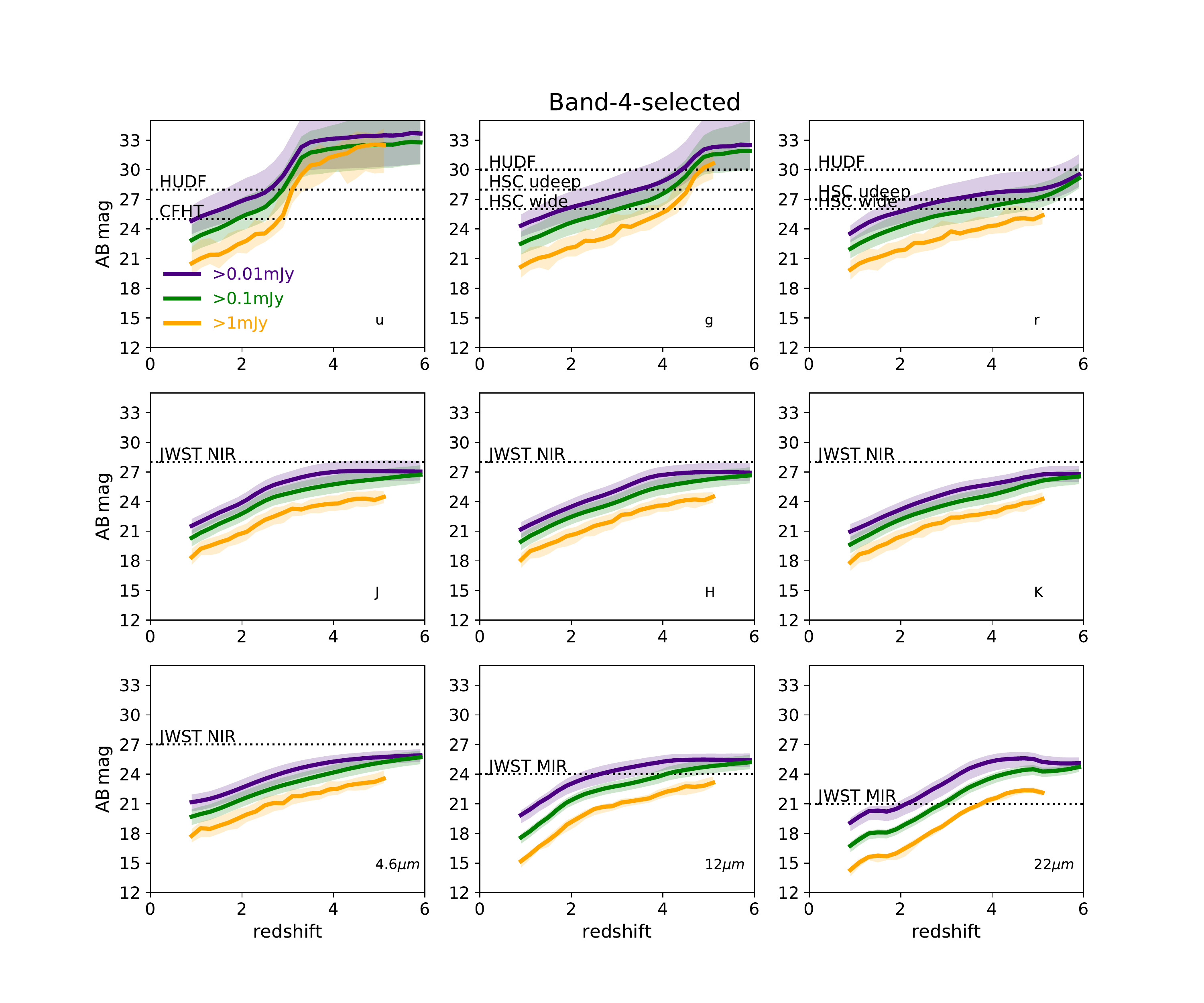}
\caption{As Fig.~\ref{ncountsband7selec} but for ALMA band-4 selected galaxies.}
\label{ncountsband4selec}
\end{center}
\end{figure*}

Fig.~\ref{ncountsband7selec} showed the expected $u$, $g$, $r$, $J$, $H$, $K$, $4.5\mu$m, $12\mu$m and $22\mu$m apparent magnitudes for band-7 selected sources as a function of redshift. Here we show the equivalent for band-6 (Fig.~\ref{ncountsband6selec}) and band-4 (Fig.~\ref{ncountsband4selec}) selected sources. 
We remind the reader that due to the wavelength range of the IR templates of \citet{Dale14} used in this work, the observer-frame band-4 emission is only well defined at $z \ge 0.84$.
At fixed flux and redshift, galaxies selected in band-4 are brighter than those in band-6 and band-7. 

\begin{figure*}
\begin{center}
\includegraphics[trim=35mm 0mm 26mm 10mm, clip,width=0.99\textwidth]{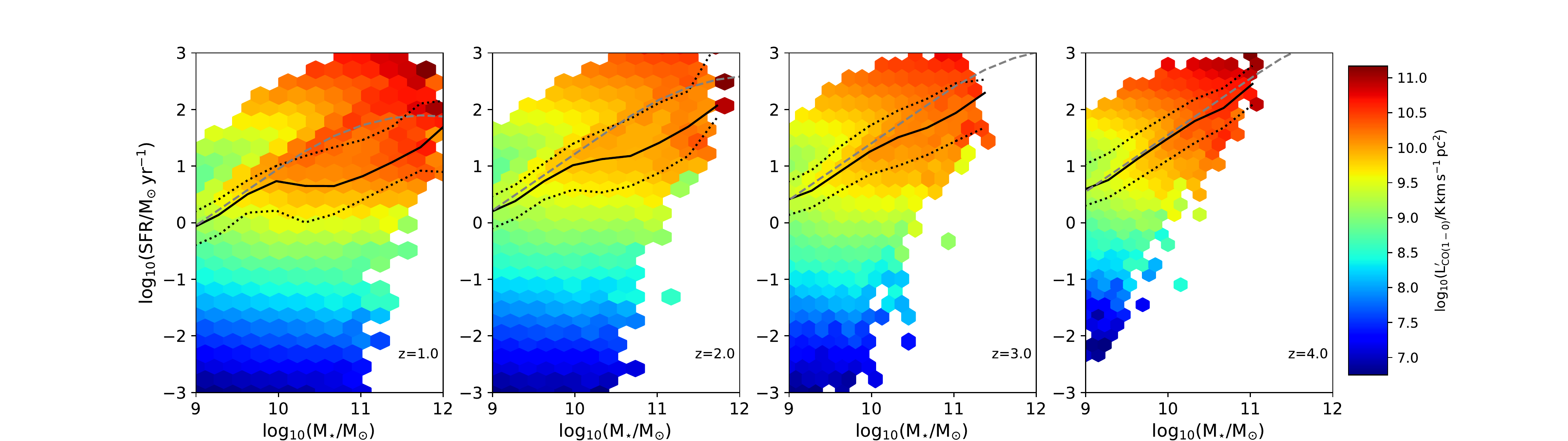}
\caption{SFR vs. stellar masses plane at four different redshifts from $z=1$ to $z=4$, as labelled, using a $\Delta z=0.15$. Pixels with $\ge 10$ galaxies are coloured by their median CO(1-0) brightness luminosity. Solid and dotted lines show the median SFR and $16^{\rm th}-84^{\rm th}$ percentile ranges in bins of stellar mass, respectively, of all galaxies with $\rm SFR>0$. For reference, we also show the MS inferred observationally by \citet{Schreiber15} as dashed lines.}
\label{tempMSCO}
\end{center}
\end{figure*}

Fig.~\ref{tempMSCO} shows the variations in the CO(1-0) brightness luminosity in the SFR-stellar mass plane in the redshift window $1\le z\le 4$. Analysis of this figure is presented in $\S$~ref{COmodel}.

\section{Spline fits of the SFR and H$_2$ density redshift evolution of SMGs}

We fit the evolution of the CSFRD and $\rho_{\rm H_2}$ of Figs.~\ref{SFRevo}~and~\ref{H2evo} of all the flux selected samples with a spline function and present the parameters in Table~\ref{TableSplineFits}. These can be compared to observational estimates as ALMA surveys become public.

\begin{table}
    \centering
        \caption{Best spline fit parameters in Figs.~\ref{SFRevo}~and~\ref{H2evo} for the $4$ different flux samples in three ALMA bands 7, 6 and 4. The equations fitted are ${\rm log}_{10}(\rho_{\rm SFR})|{\rm log_{10}}(\rho_{\rm H_2})=a_{0} + a_{1}\,z + a_{2} z^2 + a_{3} z^3$, with $\rho_{\rm SFR}$ and $\rho_{\rm H_2}$ being in units of $\rm M_{\odot}\, yr^{-1}\,cMpc^{-3}$ and $\rm M_{\odot}\, cMpc^{-3}$, respectively.}\label{TableSplineFits}
    \begin{tabular}{c|c|c|c|c}
        \hline
        Sample & $a_{0}$ & $a_{1}$ & $a_2$ & $a_3$ \\
        \hline
        \hline
        SFR & Evolution & & &\\
        \hline
        \hline
        band 7 & & & &\\
        \hline
        $S<0.01\,\rm mJy$ & -2.93 & 0.4 & -0.06 & 0.0015\\
        $0.01<S<0.1$~mJy &  -2.76 & 0.38 & -0.06 & 0.0014\\
        $0.1<S<1$~mJy & -2.1 & 0.29 & -0.06 & 0.0002\\
        $S>1$~mJy & -2.1 & 0.46 & -0.12 & $-10^{-7}$\\
        \hline
        band 6 & & & &\\
        \hline
        $S<0.01\,\rm mJy$ &  -2.76 & 0.36 & -0.06& 0.001\\
        $0.01<S<0.1$~mJy &  -2.4 & 0.25 & -0.05& 0.001\\
        $0.1<S<1$~mJy & -1.95 & 0.23 & -0.05& -0.0003\\
        $S>1$~mJy & -2.38 & 0.49 & -0.1& -0.001\\
        \hline
        band 4 & & & & \\
        \hline
        $S<0.01\,\rm mJy$ & -1.9 & 0.14& -0.003& $4\times 10^{-4}$\\
        $0.01<S<0.1$~mJy &  -1.76 & 0.09& -0.003& $6\times 10^{-4}$\\
        $0.1<S<1$~mJy & -2.6 & 0.56& -0.12& 0.004\\
        $S>1$~mJy & -3.17 & 0.19& -0.1 & 0.006\\
        \hline
        \hline
        H$_2$ & Evolution & & &\\
        \hline
        \hline
        band 7 & & & &\\
        \hline
       $S<0.01\,\rm mJy$ & 6.1 & 0.4& -0.05& $8.5\times 10^{-4}$\\
        $0.01<S<0.1$~mJy & 6.13 & 0.5& -0.1& $4.4 \times 10^{-3}$\\
       $0.1<S<1$~mJy & 6.93& 0.2 & -0.09& $2.8\times 10^{-3}$\\
        $S>1$~mJy & 6.71& 0.1& -0.06& $-3\times 10^{-3}$\\
        \hline
        band 6 & & & &\\
        \hline
        $S<0.01\,\rm mJy$ & 6.23 & 0.34& -0.05& $8\times10^{-4}$\\
        $0.01<S<0.1$~mJy & 6.5 & 0.34& -0.09& 0.003\\
        $0.1<S<1$~mJy & 7& 0.1& -0.06& 0.001\\
        $S>1$~mJy & 6.2& 0.16& -0.04& -0.006\\
        \hline
        band 4 & & & & \\
        \hline
        $S<0.01\,\rm mJy$ &  7.03& 0.096& -0.02& $-2\times10^{-4}$\\
        $0.01<S<0.1$~mJy &  7.12 & 0.002& -0.04& 0.001\\
        $0.1<S<1$~mJy & 5.7& 0.5& -0.1& 0.004\\
       $S>1$~mJy & 4.8& 0.3& -0.15& 0.01\\
        \hline      
    \end{tabular}
    \label{tab:my_label}
\end{table}

\section{Intrinsic properties of galaxies in the $\rm (u-r)$ vs. $\rm (r-J)$ colour plane}\label{AppColsSMGs}

{Table~\ref{TableIntrinsicPropsCols} shows the median stellar mass, sSFR, distance to the MS and dust-to-stellar mass ratio of galaxies with $S_{\rm band-7}>1$~mJy that fall inside/outside the passive region shown in Fig.~\ref{smgcolours}. We show this for 3 redshift bins, from $z\approx 2$ to $z\approx 4$. 
\shark\ predicts that the typical star-forming galaxies that fall in the passive region of the $\rm \rm (u-r)$ vs. $\rm (r-J)$ colour plane (and therefore could be considered contaminants) are higher stellar mass, higher dust-to-stellar mass ratio, but lower sSFR than galaxies that fall in the SF region of the colour-colour plane, at fixed redshift. The latter tend to be above main sequence, while the former tend to be below. However, given the typical criterion to classify galaxies as ``main sequence'' ($\Delta_{\rm MS}={\rm log}_{10}(\rm sSFR/sSFR_{\rm MS})=[-0.6,0.6]$; \citealt{Bethermin14}), these galaxies would all be considered as such. The trends here are qualitatively the same when we analyse instead galaxies with $S_{\rm band-7}>0.1$~mJy.}

\begin{table}
    \centering
        \caption{Median stellar mass, sSFR, $\Delta_{\rm MS}$ and dust-to-stellar mass ratio of galaxies with $S_{\rm band-7}>1$~mJy that fall inside (passive) or outside (SF) the passive region shown in Fig.~\ref{smgcolours} in the $\rm (u-r)$ vs. $\rm (r-J)$ colour plane at $3$ different redshifts. Stellar masses and sSFR have units of $\rm M_{\odot}$ and $\rm Gyr^{-1}$, respectively.}
    \begin{tabular}{c|c|c|c|c}
    \hline
         Property& $\rm log_{10}(M_{\star})$ &   $\rm log_{10}(sSFR)$ & $\Delta_{\rm MS}$ & $\rm log_{10}(M_{\rm dust}/M_{\star})$ \\
        \hline
        $z=2\pm 0.5$ &  &  & & \\
        \hline
        passive &  $11$& $-0.62$ & $-0.6$ & $-2.38$\\
        SF &  $10.4$ &  $0.6$& $0.6$ & $-2.52$\\
        \hline
        \hline
        $z=3\pm 0.5$ &  &  & & \\
        \hline
        passive &  $10.7$& $-0.11$ & $-0.28$ & $-2.39$\\
        SF &  $10.2$ &  $0.79$& $0.59$ & $-2.44$\\
        \hline
        \hline
        $z=4\pm 0.5$ &  &  & & \\
        \hline
        passive &  $10.3$& $0.56$ & $0.22$ & $-2.38$\\
        SF &  $10.1$ &  $0.95$& $0.54$ & $-2.43$\\
        \hline
    \end{tabular}
    \label{TableIntrinsicPropsCols}
\end{table}

\bsp	
\label{lastpage}
\end{document}